# HIGH-CONTRAST OBSERVATIONS IN OPTICAL AND INFRARED ASTRONOMY

Ben R. Oppenheimer[1,2] and Sasha Hinkley[1,2]

[1]Department of Astrophysics, American Museum of Natural History, 79th Street at Central Park West, New York, NY 10024 USA; e-mail: bro@amnh.org, 212-313-7921, Fax: 212-769-5007

[2]Department of Astronomy, Columbia University, New York, NY 10027 USA; e-mail: shinkley@amnh.org

Short Running Title: High-Contrast Observations

Abstract

High-contrast observations in optical and infrared astronomy are defined as any observation requiring a technique to reveal a celestial object of interest that is in such close angular proximity to another source brighter by a factor of at least $10^5$ that optical effects hinder or prevent the collection of photons directly from the target of observation. This is a relatively new type of observation that enables research on previously obscured parts of the Universe. In particular, it is most applicable to Comparative Planetary Science, a field that directly attacks such questions as "how common are planetary systems? What types of planets exist, and are there planets other than Earth that are capable of supporting life as we know it?" We survey the scientific motivations for high-contrast observations, provide an overview of the techniques currently being used or developed, and discuss some ideas and studies for future prospects.

## 1. INTRODUCTION

In some sense, all of optical and infrared astronomy requires "high-contrast" observations. Indeed, the Sun irradiates the surface of the Earth with about $10^{35}$ photons per second in the wavelength span between 0.5 and 5 µm. In contrast, the full Moon's irradiation of Earth is about a million times smaller. Vega, one of the brightest stars in the sky, irradiates the Earth at a rate that is about another million times smaller, with roughly $10^{24}$ photons per second. This is $10^{-11}$ times the Earth-bound photon flux of the Sun. Beyond that, state-of-the-art, deep observations in optical astronomy have detected objects even $10^{-13}$ times fainter than Vega. Somehow, astronomers have picked one photon from such an object for every $10^{24}$ from the Sun.

Fortunately, photons travel in extremely well-determined directions, and we have a persistent natural eclipse of the Sun with half of the surface of Earth immersed in night at any given moment, vastly reducing, by about 18-20 orders of magnitude, not only the number of photons from the Sun directly incident on a ground-based telescope, but also the number entering such a telescope due to atmospheric Rayleigh scattering (and other less-important sources of sky background). Furthermore, in space there is no atmosphere and only minimal ambient dust in our solar system to scatter solar photons into a telescope. Thus, to study many of the objects in the sky, nothing more than a standard telescope (to select photons from precisely determined directions) and suitable instrumentation (to analyze those photons) is needed to study objects that are not next to the Sun's position in the sky, or that happen to be in the darkness of the night sky.

> **Contrast:** The ratio of intensity of light between a brighter and a fainter object.

Nature provides us with the "contrast" we need to study of much of the universe.

Imagine, however, attempting to study Vega when it is just 0.1 arcseconds off the limb of the Sun. Somehow one must filter the light of the Sun from that of Vega. In fact, during the famous solar eclipse of 1919, several bright stars in the Hyades were photographed within a few arcseconds of the Sun's limb, confirming the prediction of general relativity, in one of the most important observations of the 20$^{th}$ century, that the apparent positions of these stars would be distorted by almost 2 arcseconds due to the gravitational influence of the Sun (Dyson *et al.* 1920). These observations, though, required the eclipse, which allowed the stars to shine more brightly than the background of light due to the solar corona and atmospheric scattering. In truth, these stars were at least $10^{12}$ times fainter than, and within a few arcseconds of, the Sun. These observations, along with Lyot's (1939) coronagraphic observations of the Sun's corona, possibly qualify as the first "high-contrast" observations in optical astronomy. Close proximity and a vast difference in brightness are the critical elements of what we mean by "high contrast" for the purposes of this article. More precisely, we define "high-contrast observation" as any observation in which the object being studied is detected with another object in the field of view, that is at least $10^5$ times brighter, and which is in such close angular proximity to the target object that its light due to scattering or diffraction would prevent the observation without special conditions or methods to suppress its light.

Clearly, high-contrast observations have led to fundamental results in physics, as well as enabled fields such as observational solar physics. Furthermore, as has become increasingly clear, especially over the past two decades, there are fascinating parts of the universe, that we have only just begun to observe, because a bright object, such as a star, obscures the region of interest where objects $10^5$ to $10^{15}$ times fainter exist. These regions, the close vicinities of our stellar neighbors, and the objects in them may have important connections and clues to the origins and evolution of stars, life, the Earth, and our solar system, and may also yield answers to some of the most profound questions in this field, such as "How common is life in the universe?" or "Are planets like Earth rare?"

At this point in time, high-contrast observing is primarily used in three subfields of astronomy: comparative exoplanetary science and star and planet formation. Such types of observations can also be applied to the study of advanced stages of stellar evolution where significant outflows from aging stars are present, although little has been done in this area. In actuality, the first three areas are intrinsically linked and form what is becoming an increasingly multi-disciplinary field of research in its own right, not merely a subfield of astrophysics. Comparative exoplanetary science—the study of planets in general, not just those in our solar system, how they form and evolve, their apparent diversity and their prevalence around stars—requires input from fields as diverse as geology, physics, astronomy, chemistry and, perhaps ultimately, biology. In addition, the conduct of this research requires some of the most precise engineering and control of light ever achieved, in some cases pushing the boundaries of current technologies and therefore requiring research and development as well.

## 2. SCIENCE REQUIRING HIGH CONTRAST

High-contrast observations are extremely difficult, and only few astronomers have truly attempted them. However, there is a burgeoning field of research requiring high-contrast. According to our definition outlined in the previous section, the primary motivation comes from the study of the objects and materials in extremely close proximity to stars.  We break this into two categories, comparative planetary science and advanced stellar evolution, and we explore the principal scientific questions each area seeks to answer, along with relevant observational information found to-date.  Here we purposely do not include any kind of review of the huge body of theoretical work on these subjects, merely for the reason that this paper addresses observational issues.  We only consider observational issues that require high-contrast, the rest of the article deals with the techniques used for these observations (except where necessary to support our points) .

### 2.1. Comparative Planetary Science

Fifteen years ago brown dwarfs, objects intermediate in mass between planets and stars (Oppenheimer et al. 2000, and more recently Burgasser et al. 2007), were a purely theoretical notion, after numerous surveys had only turned up one borderline object that remained controversial and inexplicable until the L spectral class was defined (Becklin & Zuckerman 1988, Kirkpatrick et al. 1999).  In adddition, exoplanets were relegated primarily to the realm of science fiction.

In 1995 that all changed, with the near simultaneous announcements at the Cool Stars IX meeting in Italy (Pallavicini and Dupree 1996) of both a bona-fide brown dwarf companion of a nearby star (Nakajima et al. 1995, Oppenheimer et al. 1995) and a peculiar Jupiter-sized planet orbiting a Sun-like star (Mayor and Queloz 1995).  At present hundreds of astronomers around the world are working on substellar companions of nearby stars and brown dwarfs.  Some 500 L-type brown dwarfs are known, and nearly 100 T-dwarfs have been identified and studied spectroscopically.  About

> **Planet and Brown Dwarf**
> The definition of these terms is highly controversial.  For the purposes of this article, and to avoid digression into the details of the arguments, we use the convention of Oppenheimer et al. 2000, which splits the classes as follows: $0.075\ M_\odot > M_{BD} > 13\ M_J$ and $M_P < 13\ M_J$, where $M_{BD}$ and $M_P$ are brown dwarf and planet mass, respectively.  We do not bother with a lower mass limit for planets here, given the primary subject of this article.  This definition is based on considerations of internal physics of such objects, not on formation mechanisms.

20 of these were found as companions of stars or other brown dwarfs (e.g. Metchev & Hillenbrand 2008 and references therein; vlmbinaries.org; Burgasser et al. 2007; Burgasser, Kirkpatrick and Lowrance 2005).  Also, nearly 300 planets outside our solar system have been identified (Udry and Santos 2007).

These populations of objects, which, we suggest below, are intimately and intrinsically related, offer a vast diversity of salient properites.  This challenges the concept in astronomy that most celestial bodies can be fundamentally understood by measuring only a few basic parameters, as suggested by the Vogt-Russell theorem, whereby knowing the mass and metallicity of a star reveals the entire nature of that star, including its evolutionary path and all other fundamental parameters.  Such a simplification has less and less utility and meaning as one proceeds to lower and lower masses along the stellar main sequence.  For example, in the brown dwarf regime (below about $0.075\ M_\odot$ where $M_\odot$ is the mass of the Sun), a chemistry, far more complex than what exists in any stellar atmosphere, has tremendous effects on the emergent spectral energy density and affects the dynamics and physics of the objects themselves (Burrows et al. 2005; Baraffe et al. 2003, Oppenheimer et al. 1998, Saumon et al. 2000).  In the planet-mass regime (commonly defined as objects below roughly 13 $M_J$ where $M_J$ is the mass of Jupiter; see Side Bar), one need only take a very superficial look at the objects in our solar system to see a vast diversity. Indeed, the giant planets of our solar system are all roughly of the same radius, of nearly the same metallicity and presumably of the same age. Yet the spectra and general appearances of Jupiter, Uranus and Neptune are all quite different. An inventory of well-studied moons of the solar system (e.g. Rothery 1992) as well as the terrestrial planets, again, reveals that a few simple parameters are insufficient to understand these objects's physical and chemical structures and processes in the context of their observable features. More than that, a comprehensive theory of planet formation, evolution and constitution cannot be derived without spectroscopic and astrometric study of hundreds, or, one might hope, thousands of these objects. We have identified six main questions that must be answered to form such a theory or general understanding of planets and brown dwarfs. These questions are addressed throughout the following sections.

1. *How common are planets around stars?*
2. *Is there such a thing as "solar system architecture?"*
3. *What types of planets exist?*
4. *How do planets and planetary systems form, evolve and die?*
5. *Are there other planets capable of sustaining life as we know it?*
6. *Are brown dwarfs part of this picture or not?*

### 2.1.1 Planet Frequency and Solar System Architecture

As of late 2008, nearly 300 planets are known to orbit nearby stars. The Radial Velocity or Doppler technique (Mayor & Queloz 1995, Butler et al. 1997, Mayor et al. 2003, Marcy et al. 2005) has been the most productive method of detection, and now limited statistical studies of exoplanetary systems can be carried out. An excellent recent review of our current knowledge of exoplanets and the corresponding statistical treatment is given in Udry & Santos (2007). These surveys find that about 1% of stars have "hot jupiters" in extremely short orbits, while about 5-11% of stars roughly similar to the Sun have planets orbiting them (Udry & Santos 2007, Cumming et al. 2008). Also, 25% of higher metallicity stars ([Fe/H] > 0.3) surveyed by Fischer & Valenti (2005) have gas giant planets, while fewer than 3% of stars with -0.5 < [Fe/H] < 0.0 have planets detected by the radial velocity technique.

It is critical to note here, and we will discuss this in more detail in §2.1.5, that all of this information, representing a huge expansion in human understanding of planets since the early 1990s, comes from heavily biased surveys, and all of these numbers should be considered rough initial attempts at, perhaps even lower-limits to, the planet frequency question. This issue is addressed in significant depth in Cumming et al. (2008) and Metchev & Hillenbrand (2008). Radial velocity surveys, for example, are not sensitive to companions beyond about 5 AU, and for all practical purposes never will produce statistics at significantly wider separation. Without characterizing the exoplanetary systems on wider orbits, our knowledge of exoplanetary systems remains incomplete. At this point, a significant way radial velocity and other indirect techniques can yield additional information is through exploration of a broader range of stars, especially in monitoring the most common stars in the Universe, the M-dwarfs (Lunine et al. 2008). High contrast imaging is ideal for characterizing planets on wider orbits. Indeed, a 10-m telescope imaging at H-band (1.6 μm) has a 32 milli-arcsecond diffraction limit. Such an instrument could resolve a planet on a 5 AU orbit around a star at 150 pc, approximately the distance to the Orion star forming region.

Despite these biases, some completeness corrections can be made to attempt to reveal the underlying nature of the planet population of the Galaxy in general. Cumming et al. (2008) performed a rigorous statistical analysis for 585 stars in the Keck Planet Search concluding that 10.5% of stars have a planet in the mass range $0.3 - 10$ $M_J$ and period 2-2000 days. Extrapolating these results, they conclude that between 17% and 20% of stars possess gas giant planets within 20 AU, and about 11% of the exoplanetary systems have multiple planets. The true fraction of stars with multiple planets is likely to be significantly higher, due to the biases in the radial velocity surveys (Udry & Santos 2007). This issue may also be directly relevant to the fraction of stars with brown dwarf companions, as we discuss in §2.1.5. Forthcoming high contrast surveys will be able to confirm, or at least further constrain these fractions over a much larger range of orbital parameters.

In our own solar system, many ideas have been suggested to explain why the giant planets and terrestrial planets seem segregated in their orbits about the Sun. Indeed this notion that a simple ordering of the solar system exists predates even Bode's law, first published in 1778, that suggested the existence of then-unknown planets, which were later discovered, along with the asteroid belt, in roughly the predicted locations (Bode & Oltmanns, 1823). With observations of many other solar systems now possible, can we find patterns in orbital characteristics such as eccentricity or semi-major axis as a function of mass or other planet properties? Do all solar systems have Kuiper Belts, comets and Oort clouds?

The two parameters best determined for most exoplanets are the mass and semi-major axis of the orbit. Plotting these two parameters (as we show in Figure 1, in four different representations of the same data) reveals several striking features. First of all, there is no obvious pattern. Points seem to occupy almost any part of the parameter space that has been probed, and regions poorly surveyed are indicated.

Figure 1 (especially in the two left-hand panels) shows a clear deficiency of massive planets (>5 $M_J$) and brown dwarfs (> 13 $M_J$) in the radial range of 0.1 to 1 AU separations, a region where radial velocity surveys are very sensitive. This is a real deficiency and can be interpreted by considering migration of planets, something that suggests that a generalized planetary system architecture is not, probably, common. For example, it may be easier for massive planets to form at larger distances from their stars, where the feeding zones during formation are larger. If these massive planets migrate to become "hot Jupiters," in the region closer than 0.1 AU, why do none of them seem to stop in the 0.1 to 1 AU region? (See also §2.1.5.)

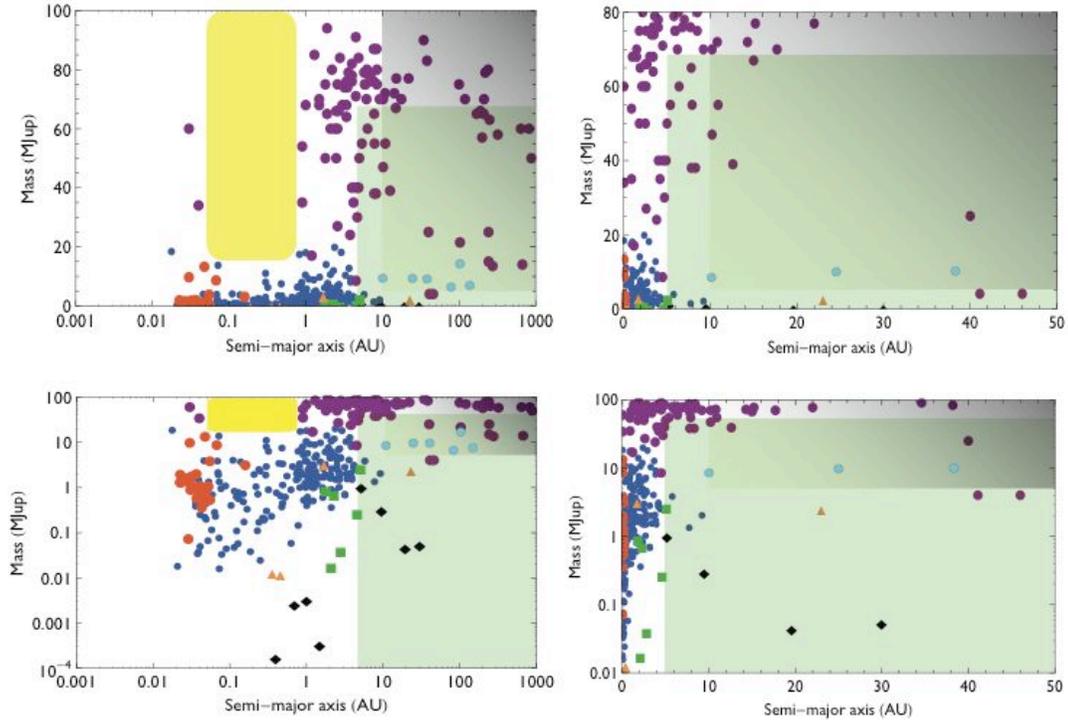

**Figure 1.** Mass vs. semi-major axis for all known objects in the mass range from $10^{-4}$ to 100 $M_J$ in orbit around nearby stars or brown dwarfs, including objects detected with the radial velocity method (blue points), the transiting technique (red), microlensing (green), pulsar timing (orange triangles), as well as the planets of our solar system (black), and objects found in direct imaging campaigns (purple circles). Data were compiled from exoplanet.eu, Burgasser (2007) and vlmbinaries.org). Data are represented in four different versions of axis scaling to emphasize different aspects of the parameter space. The yellow region indicates the only region that is clearly devoid of objects based on sensitivities of various surveys. The green region is almost entirely unobserved except in a few cases for extremely young objects (and the solar system). The grey regions show roughly where direct imaging surveys have placed some constraints on this parameter space, with darkness qualitatively representing completeness. This figure was made with generous help from R. Soummer.

Examining the distribution of planets versus the semi-major axis alone, regardless of mass, for the radial velocity planets is instructive by itself. The semi-major axis distribution for radial velocity discovered planets shows a distinctly bimodal shape—peaking at an orbital period of ~3 days and ~1000 days (See Udry & Santos 2007, especially their Figure 4). The peak near three days is likely a result of a migration process. The peak towards 1000 days (~2 AUs) may not, in fact, be a peak at all. The long observational periods for separations like this render the region beyond a few AU incomplete to the radial velocity surveys. However, even a flat extrapolation of this distribution suggests an abundance of planets at 5-20AU, which would double the rate of occurrence of planets (Marcy et al. 2005). High contrast imaging surveys are critical for constraining this occurrence rate, and probe the green areas in Figure 1.

Apart from low-mass companions of stars, a number of other related high contrast results can help fill in the picture of the close vicinities of stars. Several circumstellar rings and disks have been imaged and a few have been studied spectroscopically. These observations suggest that structures like our own Kuiper Belt may in fact exist around other stars. Notably, Figure 2 shows two such rings, imaged around the star Fomalhaut and HR 4796A (Kalas et al. 2008, Kalas, Graham & Clampin 2005, Schneider et al. 1999), both of which have been directly compared to the Kuiper belt.

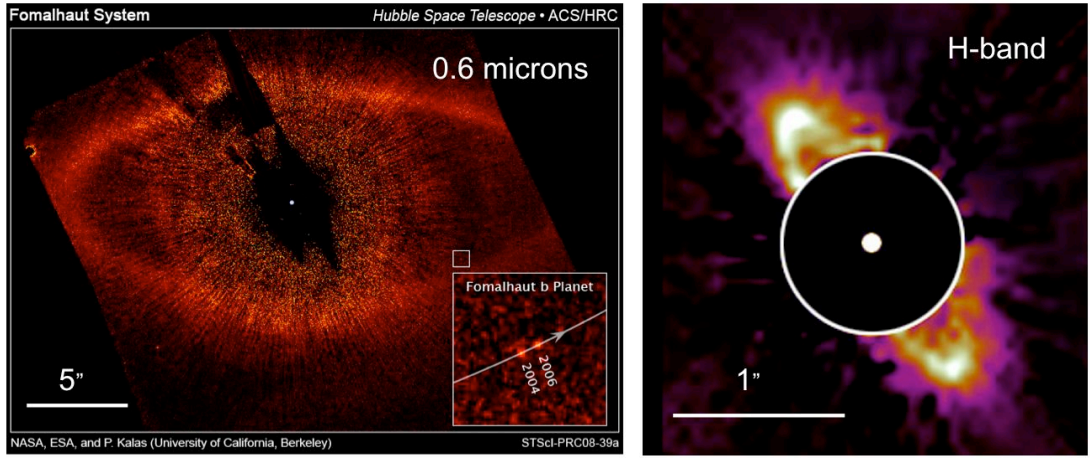

**Figure 2**. Images of the ring of debris around Fomalhaut (left; Kalas et al. 2008, courtesy NASA, ESA, P. Kalas, J. Graham, M. Clampin) and HR 4796A (right; Schneider et al. 1999, Courtesy B. Smith, G. Schneider and NASA).

The bottom line is that there is insufficient information on planetary systems in general to see whether any sort of pattern, like the architecture of our solar system, is prevalent. If anything, the data suggest that there are only weak indications of a standard architecture. These systems need to be studied on a much vaster range of companion masses and separations, something only high-contrast direct observations can achieve.

### 2.1.2 Planet Diversity

Our own solar system has a great diversity of types of planets and moons. Is this diversity mirrored in other solar systems, or will we find an even greater diversity? What, for example, would a planet five times the mass of Earth look like and how would it evolve?

Indeed, the most striking finding of the last 15 years is the *diversity* in the properties of the exoplanets found. With the discovery of the so-called "hot jupiters" (Mayor & Queloz 1995), massive planets on roughly four to ten day orbits (semi-major axes less than about 0.1 AU), scientists quickly realized that many solar systems yet to be discovered look nothing like our own. Furthermore, the mass ranges of these planets suggest no obvious classes based solely on mass. Most notably, the distribution of planet masses seems to rise sharply toward the lower masses, with a long, decreasing tail into the larger masses (>15 $M_J$), suggesting a large population of low mass (<5 $M_J$) planets yet to be discovered. Also, the distribution of exoplanet eccentricities is one of the biggest remaining mysteries in the field, and theorists have been largely unable to reproduce the distribution numerically or analytically. The eccentricity distribution (See Udry and Santos 2007, especially their Figure 6) is significantly different from the planets in our own solar system.

Marois et al. (2008) have imaged three planets in orbit around the star HR8799 using adaptive optics and angular differential imaging (ADI see §6.1, Marois et al. 2006). The three planets may be in nearly circular, face-on orbits, have projected separations of 24, 38 and 68 AU, and masses are estimated to be between 5 and 13 Jupiter masses. The authors suggest that this system is a "scaled up" version of our own solar system, yet nothing like these exist in our system. Nearly concurrently, observations by Kalas et al. (2008) of the star Fomalhaut reveal a co-moving companion at 119 AU. Dynamical models suggest an upper limit of the object of at most 3 Jupiter masses, however, a surprising lack of flux in infrared wavelengths suggest the object detected may actually be some kind of vast circumplanetary disk. These along with the direct observations of disk structures are indicators of the further diversity that high-contrast observations are bound to yield.

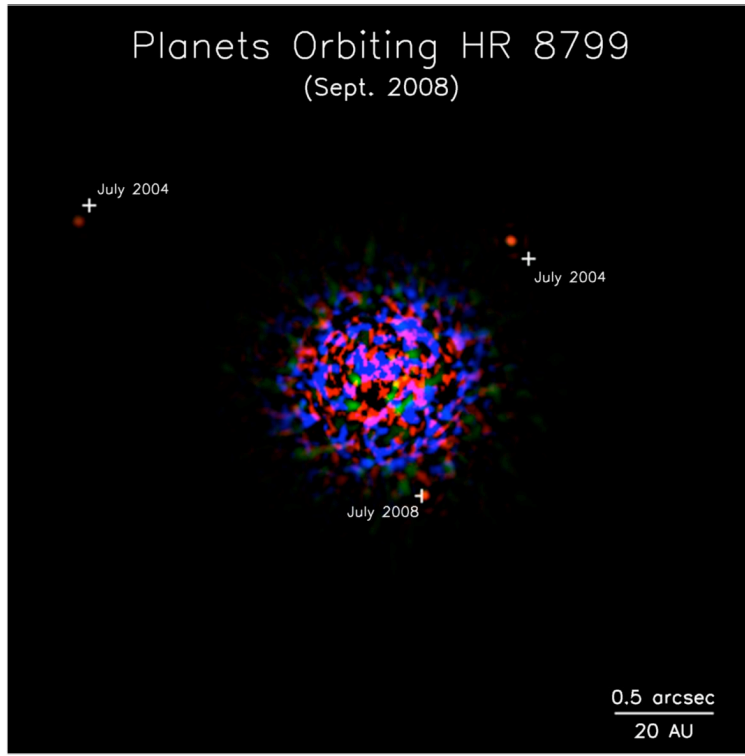

**Figure 3:** A high contrast image of the three planets in orbit around HR 8799, showing their orbital motion (Courtesy Bruce Macintosh).

The transit method (Charbonneau et al. 2000, Henry et al. 2000, Brown et al. 2001, Udalski et al. 2002) combined with radial velocity follow-up measurements, can provide the radii and thus the densities of some planets (Figure 4). In addition to the extremely diverse range of densities these objects possess, Figure 4 indicates that the size of planets, brown dwarfs, and even the lowest mass stars, are very similar between 0.001 $M_\odot$ and 0.1 $M_\odot$ as predicted by theory (e.g. Burrows & Liebert 1993). In addition to their comparable sizes, cooling theory shows that for the first 100 Myr or so, Jupiter mass planets are as hot as many brown dwarfs. Following the relation $L \sim R^2 T^4$, this implies these objects have roughly the same luminosity. Observationally this is very important because the young objects are, thus, categorically detectable with existing telescope apertures. In other words, telescope diameter is not a significant concern for detection if the starlight can be sufficiently removed from an image (See §3).

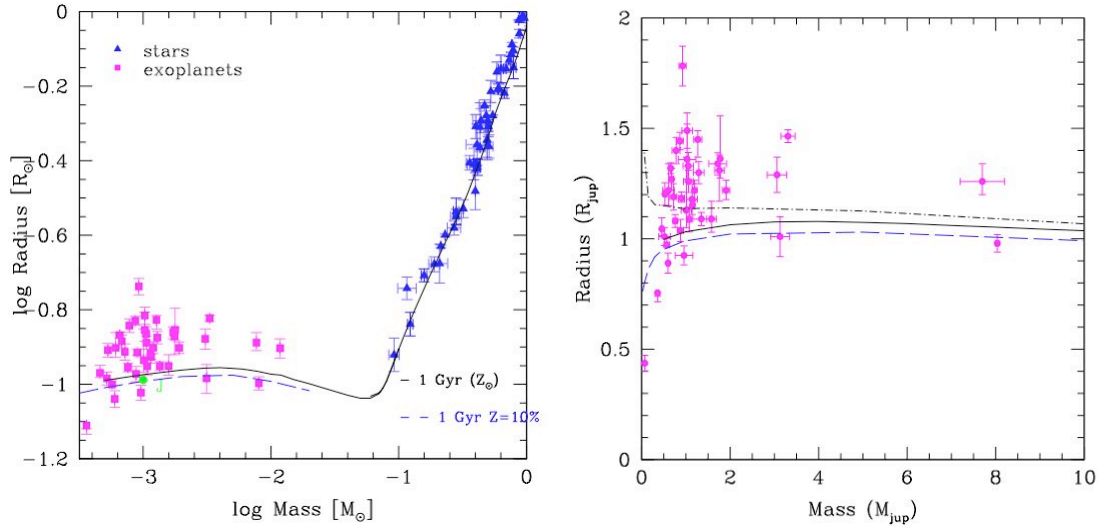

**Figure 4.** Masses and Radii for the exoplanets and stars where both values have been measured. Data is from the compendium of results by Frederick Pont (www.inscience.ch/transits) and models are from Baraffe et al. (2003, 2008). The green point in the left plot is Jupiter. Left: models are non-irradiated with solar metallicity for an age of 1 Gyr. The blue dashed line shows the relation for models with 10% heavy element enrichment at the same age. Right: Three models are shown, all for an age of 1 Gyr. The solid line is for solar metallicity non-irradiated planets, the blue dashed line for 10% metal enrichment, also non-irradiated, and the dot-dash line is for solar metallicity but with the planet at 0.045 AU from a Sun-like star. (Created by and presented here courtesy of I. Baraffe, personal communication.)

Recently, numerous mid-infrared spectra obtained during the primary and secondary eclipses of transiting exoplanets have revealed some basic characteristics of the planets, such as vertical temperature profiles, composition, and levels of atmospheric circulation (Knutson et al. 2009a, 2009b, Charbonneau et al. 2008). For HD 189733b, Swain et al. (2008) find a spectrum consistent with an atmosphere containing both water and methane (Figure 5). Also, more recent findings of planet spectra suggest that hot Jupiters may be able to be divided into two classes, with and without thermal inversion layers, depending on the degree of stellar insolation: the two objects that have the most spectral information, HD 209458b and HD 189733b, have spectra that are consistent with two quite different models. HD209458b shows evidence for an atmosphere containing a temperature inversion, while the spectrum of HD 189733b is consistent with an atmosphere absent of a temperature inversion (Charbonneau et al. 2008). However, all the transiting objects have relatively strong levels of irradiation, and high contrast imaging will allow observers to obtain spectra of a wide range of planet environments. A general picture of planetary spectra, and their diversity, requires high-contrast observations.

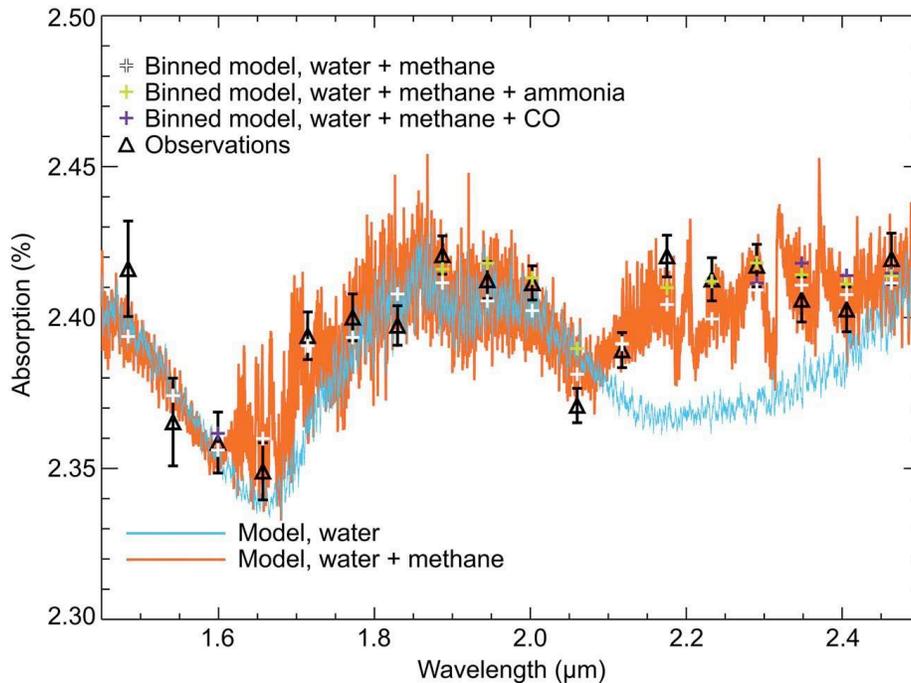

**Figure 5**. Spectrum of the planet orbiting HD 189733, showing a chemistry similar to that of brown dwarfs, but also with significant differences. The highest contrast in this spectrum is $10^4$ (courtesy G. Vasisht, from Swain et al. 2008).

### 2.1.3 Habitability and Planets Harboring Life

High contrast imaging is likely to address fundamental questions in astrobiology. These questions involve investigations of conditions necessary for life itself, and whether we can identify planets that host life by identifying chemical disequilibria induced by the presence of biologically activity.

The recent book "Extrasolar Planets and Astrobiology" (Scharf 2008) is an excellent in-depth treatment of the subject, which is also significantly dealt with in the Terrestrial Planet Finder and DARWIN project reports (see §6; Levine, Shaklan & Kasting 2006, Fridlund 2008, Beichman et al. 2007). Biomarkers, one of the cornerstones of astrobiology, are defined by Kaltenegger et al. (2006) as "detectable species, or sets of species, whose presence at significant abundance strongly suggests a biological origin." Lunine et al. (2008) state that evidence for biology is "simultaneous detection of $O_2$ or $O_3$, along with a reduced gas such as $CH_4$ or $N_2O$. This is a powerful diagnostic for a disequilibrium condition" (See also Kasting & Catling (2003), Kaltenegger & Selsis (2008), and Turnbull (2006)). These biomarkers have spectral features that are even detectable with a very modest spectral resolution of 30 to 40 in the near-IR. In fact, some high contrast experiments already operate with such resolution (§5, §6).

### 2.1.4 How do planets and planetary systems form, evolve and die?

Are there specific conditions necessary for a star to host a solar system? Is the formation of a solar system really a distinct process from that of the formation of the star, or is it something that naturally emerges from the star formation process? What would a set of fifty 1 $M_J$ planets with ages spanning 0.1 to 10 Gyr all have in common? Do planets survive the final evolutionary stages of their parent stars?

Our understanding of planet formation can only be complete when a broader understanding of the physics of circumstellar disks is in place. Indeed, as Duchene (2008) points out, since disks are the birthplace of planets, a comprehensive understanding of how these systems form and evolve, may shed more light on the planet formation process than a statistical evaluation of systems with fully formed planetary systems (Udry & Santos 2007). Characterization of the dust mass, dust size distribution, and gas content all have profound impacts on the two leading theories of giant planet formation: the gravitational instability model (Boss 1997, Mayer et al. 2002), and the core accretion model (Mizuno 1980, Pollack 1996, Laughlin et al. 2005, Alibert et al. 2005).

The architecture of a circumstellar disk has previously been determined by fitting the infrared excess in a spectrum of the system, to estimate the extent of flaring, the presence of holes or gaps, and the temperature distribution through the disk (Kenyon & Hartmann 1987, Bertout, Basri & Bouvier 1988, Chiang & Goldreich 1997). High contrast imaging of disks in the optical and near-infrared can *reveal* physical parameters that are either difficult to obtain, or are hopelessly degenerate with other parameters. For instance, high angular resolution imaging reveals disk radius, inclination and asymmetries.

High contrast imaging also allows the observer to study the light being scattered from dust particles in a circumstellar disk. In particular, multiwavelength observations of disks probe different depths in the disk, since the dust opacity is highly wavelength dependent. Longer wavelength imaging probes closer to the midplane of the disk, where the density of larger particles is higher due to settling. Near-IR observations are useful for probing the surface properties of disks, especially as demonstrated by AO polarimetry in the J, H, and $K_s$ bands (Perrin et al. 2006, and references therein) for Herbig Ae stars. Also, by studying the distribution of the scattered light in a circumstellar disk, the observer can gain clues to the nature of the scattering phase function, and hence make inferences about the size of the dust grains. Studying the scattering properties in this way can directly address how dust grains grow within the disk (Fitzgerald et al. 2007, Graham et al. 2007). Each of the techniques described above allow the observer to disentangle the relevant physical parameters that may be degenerate in an approach that uses low-contrast observations of spectra with accompanying modeling of the energy distribution.

Recent high contrast observational programs have detected disk asymmetries: evidence for interaction between a companion and a disk (e.g. Oppenheimer et al. 2008, Kalas, Graham and Klampin 2005). The recent images of an unresolved object interior to the dust disk around Fomalhaut (Kalas et al. 2008 and Figure 2), verified earlier suggestions (Quillen 2006, Kalas, Graham & Klampin 2005) that a small companion was sculpting the sharp inner edge of the circumstellar ring, and causing an offset between the disk center and central star. A recent polarimetric image of the AB Aurigae circumstellar disk (Oppenheimer et al. 2008) shows features consistent with dynamical models involving a companion inducing density amplifications in the disk, near two of the presumed Lagrange points relative to the star and the putative object (Moro-Martin & Malhotra 2002, Wolf, Moro-Martin & D'Angelo 2007). The imaging data (shown in Figure 6) also reveals a point source in a clearing of the disk at ~100 AU, though it was detected with poor signal-to-noise ratio. The authors suggest that this point source may be an overdensity in the disk due to accretion onto an unseen companion or a direct detection of a 5 to 37 Jupiter-mass object. Thus there can be no question that direct imaging of disks will, in the long term, provide clues to the nature of companion formation and evolution.

It is important to note the majority of circumstellar disks have optical depths that are significantly lower than those disks that have already been imaged. Only a few solar neighborhood stars have dust optical depths as large as HR4796A (Figure 2) or β Pictoris (Smith & Terrile 1984), with $\tau \sim 10^{-3}$. As stars age, the circumstellar material surrounding them becomes increasingly more optically thin, as confirmed by observations spanning stellar lifetimes of a few Myr to a few Gyr (e.g Hillenbrand 2005, Silverstone et al. 2006, Meyer et al. 2008). Of course, every increasing age decade contains more stars. Thus there must be numerous older stars in the solar neighborhood with faint disks, hidden from view by a lack of high-contrast capable systems. As high contrast imaging matures, it will fully characterize this most common population of faint circumstellar disks.

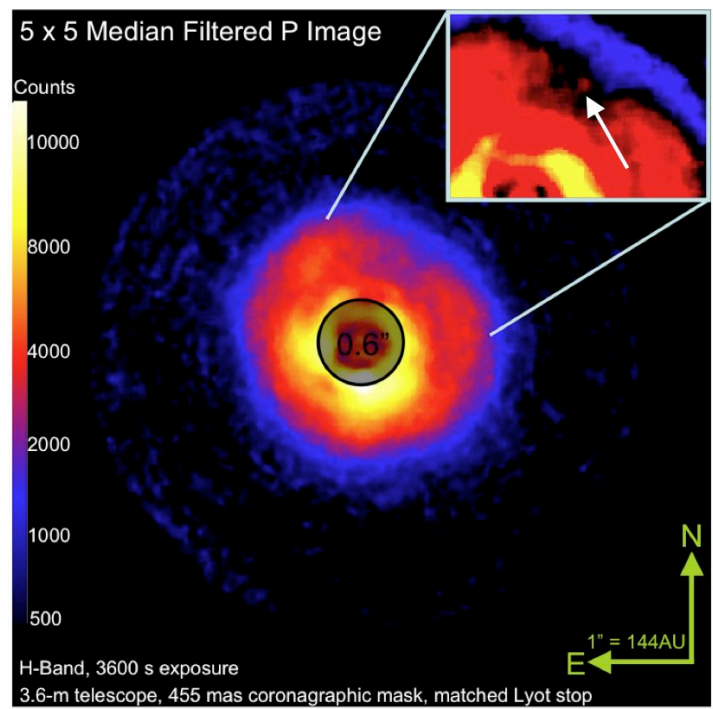

**Figure 6**. Polarimetric image of the circumstellar material around AB Aurigae, obtained in the near IR H band (1.6 μm). The structure suggests that a small body is forming at the upper right (inset with arrow) and causing amplifications of dust ahead of and behind the body in its orbit around the star (from Oppenheimer et al. 2008).

Finally we note that the evolution of planets can only be understood in detail through spectroscopy of hundreds of them at various ages. This absolutely requires high-contrast observations. Also, the evolution of these bodies, although modeled in significant detail (e.g. Burrows et al. 2005, Baraffe et al. 2003, 2008), remains a region of this field that is purely theoretical at this point. Perhaps the only observational constraint, and a weak one at that, is that planet-mass bodies have been discovered in orbit around pulsars (Wolsczan and Frail 1992). This suggests that either planets survive the "death" stages of their main sequence stars, or can be formed again out of the debris of late stellar evolution. In this context, the science mentioned in §2.2 may even be related to comparative planetary science.

### 2.1.5 How do the brown dwarfs fit in?

Most researchers in exoplanetary science assume that brown dwarfs and planets are not formed in the same way and thus distinguish the two as separate populations, generally dealt with completely independently. In fact observations contain scant, if any, evidence for such an assumption. Brown dwarf companions may provide clues regarding the interconnectedness of star and planet formation processes as an integrated process, not two separately treatable physics problems, perhaps something that even scales to the formation of disks around brown dwarfs and rings around evolved planets.

The simplest way to compare planets and brown dwarfs is to examine the mass vs. separation parameter space, as shown in Figure 1. (Note that many versions of this figure exist in the literature, but very few simultaneously plot both brown dwarf companions and planets, presumably due to the semantically induced bias mentioned above.) This figure shows the same parameter space but in 4 different versions of axis scaling. It is important to do this since the masses cover almost 6 orders of magnitude and the separations can be from sub-AU scale to 1000 AU. A cursory look at these plots shows no clear demarcation of any particular population. There is a distinct continuity of points essentially all over the parameter space, and no particular difference between the brown dwarfs and exoplanets, with perhaps one exception. In the region between 0.1 and 1 AU there are no objects above about 10-15 $M_J$ (marked yellow in the figure). Over the years various authors have described a so-called "brown dwarf desert," suggesting that brown dwarf companions of nearby stars are simply extremely rare (see for example Metchev & Hillenbrand 2008, Marcy & Butler 2000, and Grether & Lineweaver 2006). In fact, as we show in this figure this is not obviously true, except in the restricted yellow region. The rough composite sensitivity of the many surveys searching for brown dwarf companions is represented by the grey shaded region with darkness approximating the completeness of these searches in covering a statistically meaningful part of the parameter space. These sensitivities were studied in some depth by Metchev & Hillenbrand (2008), and in their Figure 13, they summarize the results of about 24 different surveys for brown dwarf companions, for which they find an aggregate, completeness-corrected brown dwarf companion rate for stars of 3 ± 3% in the separation range of 28-1200 AU. Curiously, this is entirely consistent with the few estimates of the fraction of stars with planetary systems (5-10% depending on the survey chosen). If one further considers the tremendous incompleteness of both sets of surveys, the only conclusion that can reasonably be made is that the fraction of stars with either planet or brown dwarf companions must be significantly higher. A large population of these companions remains hidden from view, and a mild deficit of brown dwarfs around 20 $M_J$ (regardless of separation) cannot be assumed to be real at this point.

Indeed, Metchev & Hillenbrand (2008) also find evidence for a universal Companion Mass Function (CMF), and state that the deficiency of wide-separation substellar companions is a natural byproduct of the CMF (shown in Figure 9), which not only differs from the field initial mass function, but also strongly suggests that a significant number of brown dwarf companions exist beyond 150 AU for stars more massive than about 0.7 $M_\odot$. The CMF and separation distribution from this study also means that (if correct), many brown dwarfs will be found around the lower mass stars at close separations (within 30 AU). Unfortunately, very few surveys (whether they be indirect or

14                                                                                                    Oppenheimer and Hinkley

direct) have studied primarily low-mass stars for companions, another area where high-contrast observations are one of the few options.

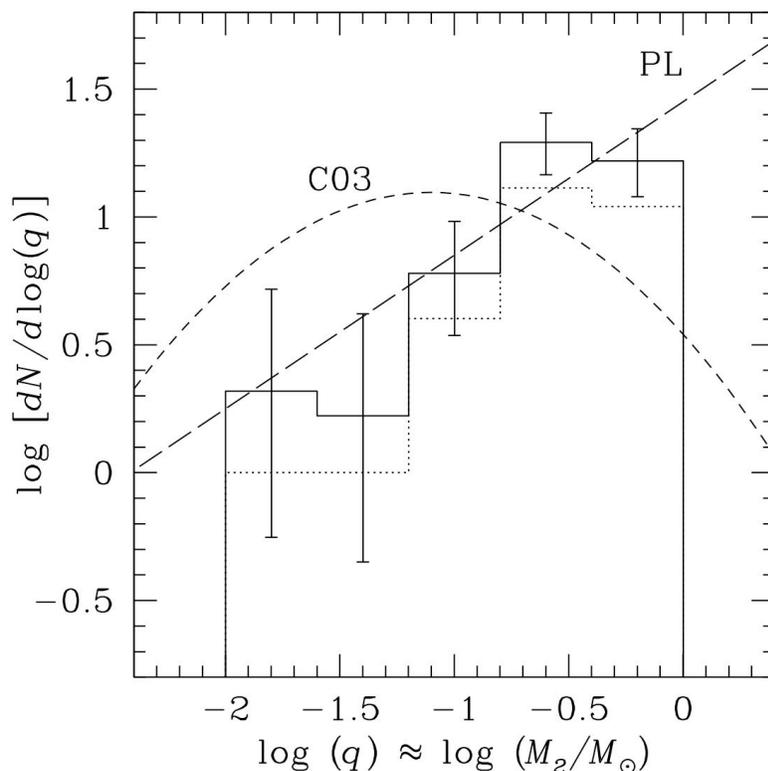

**Figure 7**. The companion mass function from Metchev & Hillenbrand (2008). Over-laid is the Chabrier (2003) initial mass function for stars, labeled C03. The solid line is completeness corrected (dotted data is raw data), and the dashed line is a power-law model fit. This function suggests that many more brown dwarfs are to be found around the lower mass stars, most of which have not been properly surveyed for companions. (Courtesy S. Metchev.)

As a final note on the issue of unprobed parameter space, the incompleteness of the radial velocity surveys is also very complex (most thoroughly addressed in Cumming et al. 2008). In addition, there may be additional biases introduced by observers whereby stars with very long-term, high amplitude variations in the radial velocity are simply removed from observation under the assumption that they are orbited by a stellar companion. This may exclude brown dwarf companions from the radial velocity searches altogether. It is certainly highly suggestive that such objects may be in the sample of stars that have been surveyed, but never were published, particularly given the distribution of purple points in Figure 2. For some unknown reason, brown dwarfs have been found as companions in direct imaging surveys, but not in the radial velocity studies, even in regions where the radial velocity studies claim significant sensitivity.

As of late 2008 several very young objects have been discovered which apparently may be brown dwarfs or may be planets. The exact nature of such objects, 2MASSWJ 1207334-393254 (Chauvin et al. 2005a); GQ Lupi B (Neuhäuser et al. 2005); and AB Pic B (Chauvin et al. 2005b), is still being debated. Regardless, these objects bear the salient properties of brown dwarfs of the L spectral class. To suggest that brown dwarfs and planets are not related ignores the facts.

### 2.2. Stellar Physics and AGNs

One area of astrophysics in which high-contrast observations have never before been applied is in the study of the late stages of stellar evolution. For example cataclysmic variables, Wolf-Rayet stars and advanced red-giant stars have potentially interesting morphology and physics in the radial range from 1 to 1000 AU. The bottom line is that this is a region of parameter space as yet unexplored in astronomy. New phenomena, or at least new data on the dynamical and chemical behavior of stellar outflows, the engine that generates heavier elements and allows already formed ones to re-enter the interstellar medium, could be obtained with suitable suppression of the light of the star to reveal these complex regions of space. However, very few Wolf-Rayet stars are bright enough for adaptive optics systems to guide on them, and so this area of stellar astrophysics has not yet shared the benefits of high-contrast techniques.

In fact, there are hundreds of cataclysmic variables within 200 pc of the Sun (e.g. Downes et al. 2001), so a relevant angular resolution required to probe the 10-1000 AU region is at worst 50 mas, within reach of existing telescopes. For Wolf-Rayet stars, the current census has the closest at about 400 pc, but more than 300 are known within the 500-3000 pc distance range (van der Hucht 2001). As such, an observation with a resolution of 50 mas could still probe a 50 AU scale (comparable to our solar system) around many such stars. The red giant census has many more objects closer to the sun, even within 50 pc (Lepine 2005).

Although study of Active Galactic Nuclei (AGNs) often involves significant brightness differences between the nucleus and the host galaxy, these objects do not explicity fit under our definition of high-contrast observations. The necessary levels of contrast are typically 10-100 (see Table 1), with a highest contrast example approaching $10^3$ (Magain et al. 2005, Floyd et al. 2004).

### 2.3. The Prototypical High-Contrast Observation

We can examine a previous set of observations as an example of how one might wish to conduct high-contrast science into the future. The discovery and detailed study of the brown dwarf Gliese 229B, in a sense, exemplifies the issue of finding and studying a faint companion of a nearby star (Figure 8). In this case, the companion is so well separated (> 7 arcseconds) and has a relatively low contrast of about $10^4$ in the optical and near IR, that when it was detected, basic stellar coronagraphs and standard spectrographs were sufficient to acquire data (Oppenheimer 1999, Oppenheimer et al. 2001). The real issue today is how this can be done for contrasts many orders of magnitude higher and for objects that are much less than an arcsecond away from their parent stars: Imagine an object $10^4$ times fainter than the brown dwarf in Figure 8 and situated at the 1 AU orbit ellipse drawn on the left-hand image. Without special techniques to remove the contaminating starlight, the observations would be impossible.

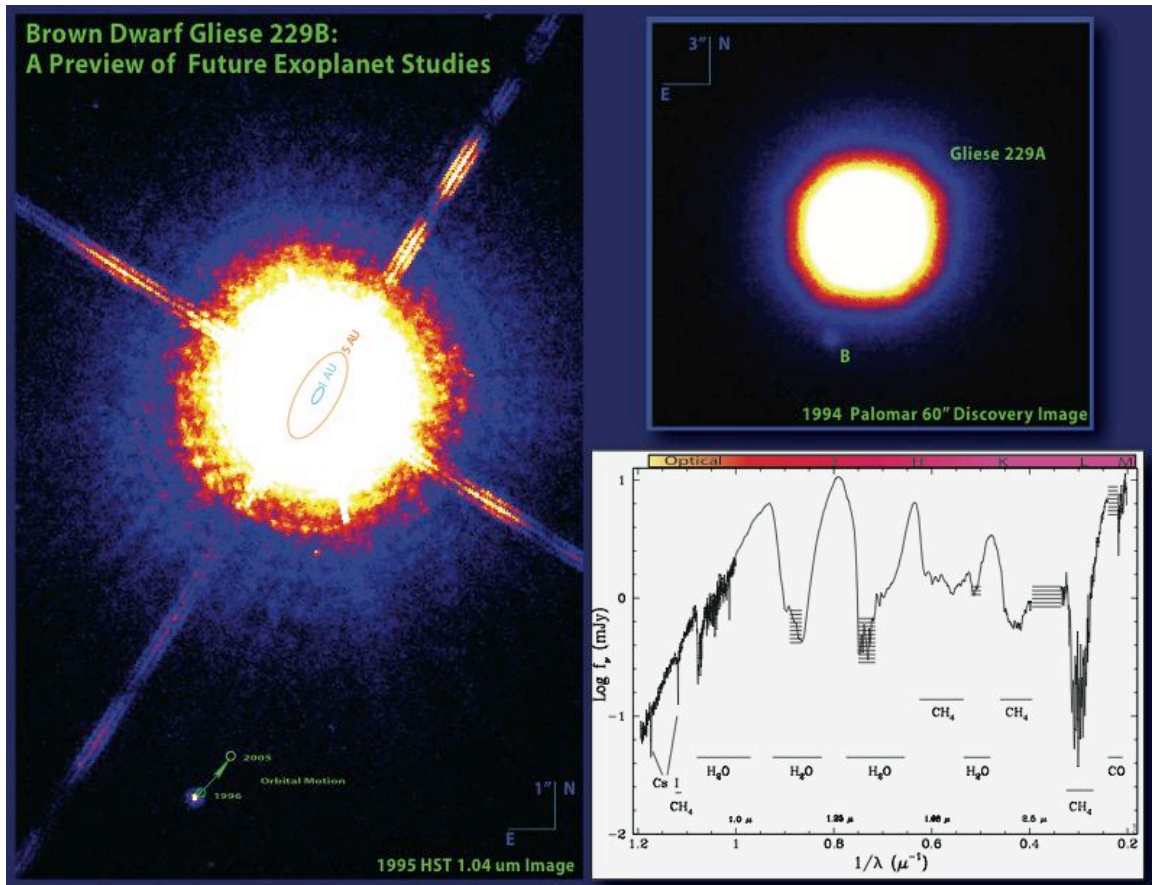

**Figure 8.** Primary data on the Gliese 229 system, an M-dwarf, T-dwarf pair discovered with coronagraphy, followed up with numerous observations to obtain orbital information and spectroscopy over a broad wavelength region. Future exoplanet and circumstellar material studies may be similar in terms of method of data collection, whereby the primary star's light is extremely well-separated, intentionally or "naturally," from the companion (at lower left in the two images above), allowing for precision astrometry, photometry and spectroscopy without contamination.

## 3. OBSERVING REQUIREMENTS AND GOALS

As we have outlined, high contrast imaging depends on control of the starlight. The science does not require significantly greater telescope apertures or a revolution in telescope technology. Rather, the high-contrast sensitivity depends on controlling the residual starlight. The rich science outlined in §2 has a very bright future. Indeed, with at least 300 planets indirectly detected, only a few with low-signal spectra measured, a handfull with radii derived, and several other key parameters determined, we have barely begun to probe the nature of this population. In addition, as the mass distribution suggests (e.g. Udry &

> **Point Spread Function**: the probability distribution function describing where photons from an infinitely distant point source of light will be detected in the image plane of an optical system. For a circular imaging device this is an Airy function, with an approximate core width (diameter at first null) of $2.44\lambda/D$, where D is the diameter of the input aperture.

Santos 2007), the majority of planets may be well beyond current detection capabilities. This is a field that is heavily driving new techniques, and thus the subject of this paper.

Estimates, based on cooling models for planets (Figure 11; Burrows et al. 2005) suggest that detailed spectroscopic study of even young (~100 Myr) and hot (~800 K) planets of about 1 $M_J$ requires a contrast of $10^8$. This depends upon the age and physical distance between the star and planet, but the point is that although one could image such planets while only distinguishing one photon for every $10^6$ from the star, probing the depths of the spectral features, with which one may study the atmospheric chemistry and physics, requires *at least* two more orders of magnitude in precision. The majority of planets are not hot and young, however, and in those cases the star will emit more than $10^{12}$ photons for every one emergent from the planet in the optical and near infrared absorption features (Figure 11). Study of planets similar to the Earth may require an additional two orders of magnitude of contrast, $10^{14}$. (See, for example, Levine, Shaklan & Kasting 2006, Traub et al. 2006).

To grasp the delicacy of such measurements, one can use an analogy. For example, taking a broad-band image of a 1 Gyr, 1 $M_J$ exoplanet is similar to taking a single picture of the Empire State Building (443.2 m high, representing the stellar point spread function; hereafter PSF, see definition box) that contains the entire building's structure but also resolves a bump in the sidewalk (representing the planet's PSF) that is only 4 μm high. Measuring the depths of the planet's absorption features would be akin to using that image to tell that the bump is 4.43 ± 0.01 μm. This would require a camera with, at the very least $10^7 \times 10^8 = 10^{15}$ pixels, along with optics that could provide both the field of view and the angular resolution to do this. Professional digital photography typically deals with arrays with a few times $10^3$ pixels on a side, and the largest scientific arrays are still at least 4 to 5 orders of magnitude smaller than those required to conduct such a silly exercise.

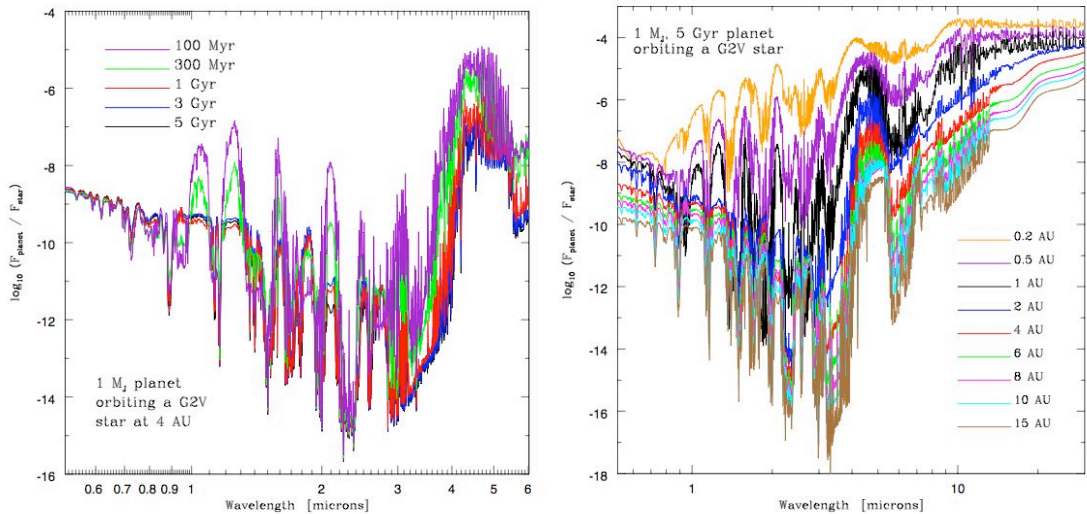

**Figure 9.** Model emergent spectra for 1 $M_J$ planets as a function of age (left) and as a function of orbital radius. The models include irradiance from the central star (chosen to be identical to the Sun), and are expressed in units of the inverse of contrast (the ratio of planet flux to star flux). (Left from Burrows et al. 2004, right from Burrows 2005.)

Analogies, though useful for appreciating the scale of a problem like this, are never perfect, and, in the case of exoplanetary science, angular resolution is not the primary problem. It merely serves as a proxy for intensity in this analogy. Indeed, even modest sized telescopes (with diameters of about 4 m) can resolve physical scales corresponding to a few AU around stars within 25 pc of the Sun. Rather, the field has been dominated by

> **Resolution:** independent sources in a perfect, diffraction-limited image can be discerned when they are separated by at least $\lambda/D$ in angle. This is also called the Nyquist criterion most commonly derived with a Fourier optics approximation. This concept can also be derived based on the Heisenberg uncertainty principle.

the development of instrumentation designed to produce high resolution images in which as many of the photons from the star are eliminated while leaving those from the planet intact. This requires understanding the physics and behavior of star light with real optics, as well as an ability to control optical and near-IR wave fronts to unprecedented precision.

In Table 1 we summarize rough estimates of contrasts and relevant angular separations that must be probed to achieve the various types of science discussed above. We also include, in this table, a few areas that were not addressed above because they do not meet the strict definition of high-contrast used herein, but they provide a comparative frame of reference. Angular resolution in the case of the exoplanet studies is calculated for a mean distance of 10 pc for the sample of surveyed stars. Note that at $\lambda = 1$ μm, an 8-m diameter circular telescope has a diffraction-limited resolution of 25.7 mas, exceeding the angular resolution requirement for essentially all of the cases in Table 1. In addition, it is important to note, as Figure 4 and our discussion in section 2.1.2 showed, telescope aperture is not a critical part of this problem for many science targets, because in the absence of the star, they (for example relatively young Juipter-mass planets) would be bright enough to detect even with modest-sized telescopes. Large apertures are only required for the older and much cooler targets of study.

## 4. THE OPTICAL PROBLEM

### 4.1. Diffracted Light

As mentioned above, the primary issue is residual starlight that contaminates and obscures the region of interest. This contamination comes in two primary forms: light due to optical diffraction, and the far more insidious distribution of light due to imperfect optics, and propagation of the light through a turbulent medium (also known, in general, as residual wave front error, and treated in §3.3).

Ideally, one would like simply to image a nearby star and delete the star itself, leaving visible whatever is in close proximity. In fact, even with a perfect, diffraction-limited imaging device, the PSF (see definition) of a telescope with a circular aperture is an Airy function, placing significant light from the star at the location of objects in close proximity to it. Indeed, at just four image resolution elements (see definition) away from the core, a region of great scientific interest (Table 1), the light from the star is as bright as $3 \times 10^{-4}$ times the central peak brightness. Thus, any object with a contrast of a few thousand or more will be fainter than the background of light due to the star. This, of course, does not mean that it cannot be detected, but for the cases outlined in Table 1, some of which would require large integration times even in the absence of the primary star, it is clear that some kind of suppression of the starlight is necessary to meet, in any practical way, the goals of §2.

For a perfect image, limited only by diffraction due to the optical system, where the PSF is known to an extremely high precision, one could simply subtract the known PSF from the image and reveal whatever is deviant from it. This could, in principle, work for any contrast range, given perfect detectors as well. However, such observations, with real detectors, are intrinsically highly inefficient, spending the majority of their time measuring the light of the star, which is of no interest based on the science goals, and which is at least $10^5$ times brighter. This results in ridiculously long integration times simply to get enough signal on the PSF to detect a tiny ($<10^{-5}$) "defect" on it due to another celestial object. Thus, even in the perfect case, one must suppress the light of the star. Most of the starlight suppression techniques outlined below mainly affect the perfect, diffraction-limited PSF, thus reducing the exposure time needed to measure such PSFs at the required precision for a given contrast.

There are two classes of techniques that attempt to remove diffracted light in high contrast observing: coronagraphy and interferometry, described in §5, after we treat the primary optical obstacle to high-contrast observations: speckles.

### 4.2. Speckles

In practice, no imaging system is perfect, with the incoming wave front of light deviating from the expected perfectly flat one by small "wave front errors." One must rely on an engineering mindset: reach a level of perfection that along with a given diffracted-light suppression technique, achieves the contrast necessary, while permitting unknown imperfections at an inconsequential level.

As an example, image subtraction, without any other starlight suppression technique, has only yielded significant results where the image quality—characterized by a "Strehl ratio," S—is above 70% (see the Hubble images of circumstellar disks, e.g. Figure 2). However, even for such high values of S (n.b. the vast majority of astronomical images have S < 1%), these image-subtraction results also clearly show that achieving high contrast at angular separations of a few λ/D is essentially impossible. The residual PSF differences from one's perfect estimate, or even a measured PSF taken moments later, are so large as to preclude the contrast of $10^5$ or better that we seek. One must control or precisely measure these tiny differences in PSF, differences that have spatial and temporal dependence that are generally not predictable. In an image with S = 70%, 30% of the starlight is distributed over the field in an extremely complex pattern of so-called speckles (e.g., Figure 10). If that light could be controlled and put into the diffracted PSF pattern, where it ought to be in a perfect imaging system, one would achieve a value of S = 100%. Speckles are the biggest challenge for any high-contrast observer.

> **Strehl ratio**: ratio of the peak intensity in a real image to that of a perfect image made with the same imaging system's fundamental parameters. Also approximated by $S \propto \exp(-\sigma^2)$ where $\sigma$ is the root mean square of the wave front error in radians, when $\sigma \ll 1$.

Before we discuss controlling speckles, it is useful to understand them conceptually, and to estimate how well they need to be controlled to permit a certain level of contrast. We use a simple approach (also described in Stapelfeldt et al. 2006) with a Fourier approximation of optics in the far field. Consider a wave front of light from a star so distant that the optical system observing it cannot resolve it. This wave front, in a perfect situation, would be a segment of a sphere centered on the star, and for all purposes extremely well-approximated by a perfect plane at the entrance pupil of the telescope. (Indeed, for a star at 10 pc, the deviation of this segment of the sphere is less than one part in $10^{17}$ for a 4-m telescope.) The telescope acts simply as a Fourier transform on this plane multiplied by the pupil shape, when it forms an image. This image, the Airy function PSF, as described in the previous section, will be modified if there is any deviation from the perfect plane at the entrance aperture. As with functional analysis, one can represent any complex function perturbed from a plane as the sum of Fourier component perturbations. Thus, consider a simple case, in which a single component, a sinusoidal ripple of small amplitude in phase is imposed on this plane. Such a ripple results in Fourier components in the image plane that appear to be fainter copies of the primary Airy function PSF, but displaced from the primary PSF by an angle which is related to the inverse of the sinusoidal wave's frequency across the pupil of the telescope. One will appear on either side of the primary PSF, resulting in a symmetric set of fainter spots, spots of exactly the size and shape that a real faint companion would have if one were at that location. Given a complicated input wave front, one can imagine thousands of such spots at essentially arbitrary locations throughout the image, each with different intensity (e.g. Figure 10), and any minor change in the wave front will move these speckles around in the image plane. The positions of the speckles are a function of the wave front perturbations and the wavelength of light observed. In addition to this, tiny amplitude perturbations in the incoming wave front's electric field also translate into speckles in the image plane (with anti-symmetric behavior), though these are generally much fainter than the ones due to phase errors. The result is a highly variable, non-smooth background against which one seeks to find very faint objects. There are numerous publications dealing with speckles. An early identification of the significance of speckle noise as a major problem in high-contrast imaging appeared in Racine et al. (1999). However, later studies (Boccaletti et al. 2003, 2004, Marois et al. 2005, Hinkley et al. 2007) began to quanitify their affect on real data.

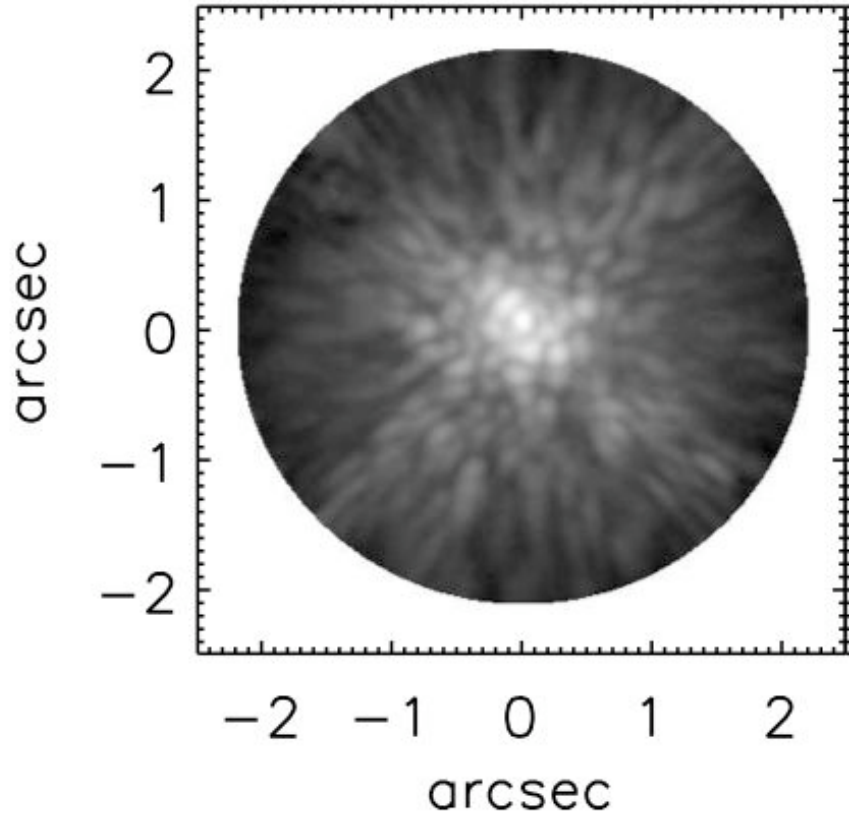

**Figure 10.** Coronagraphic image of a nearby star demonstrating the prevalence of speckles even in extremely high fidelity images. This H-band image is from the Lyot Project (Hinkley et al. 2007, Oppenheimer et al. 2003). The unocculted image has a PSF with S ~ 85% (Oppenheimer et al. 2003).

**Dynamic Range Curve**: the 5-σ detection limits expressed as the difference in magnitude between the primary star and whatever faint companion is being sought as a function of the radial separation between the two in the image.

Generally, speckles behave in such a manner that they do not obey Poisson statistics, and they represent a noise source several orders of magnitude larger than the shot-noise behavior of the underlying perfect PSF (Racine et al. 1999, esp. their Figure 2). Speckles are also highly evanescent. They can be due to changes in the turbulent atmosphere (with a timescale of milliseconds) and due to static, or semi-static features in the optics used to process the light. They exhibit a correlated noise behavior (Soummer et al. 2007), and as such, one cannot simply let them "average" out into a smooth background against which one can pick out a much fainter source (as one does for objects fainter than the uniform sky background in many deep astronomical observations). Instead, several authors have now noted that simple augmentation of exposure time to achieve a higher signal-to-noise ratio on an object in the midst of a field of speckles results in *no* increase in sensitivity once the speckle lifetime is exceeded (Hinkley et al. 2007, Duchene 2008). Such speckle lifetimes vary from site to site and with different adaptive optics systems and instrument configurations, but they tend to be on the order of a few milliseconds to ten seconds with some lasting many minutes at 0.5 <

λ < 2.5 μm (e.g. Figure 11, right).

Another aspect of this speckle noise is that it seems largely independent of the size of the telescope aperture. In Figure 11, we show so-called dynamic range curves (see definition) for systems operating at the speckle noise limit but on telescopes from 3.6 to 10-m in diameter. There is little, if any, improvement by conducting these sorts of observations on a larger telescope.

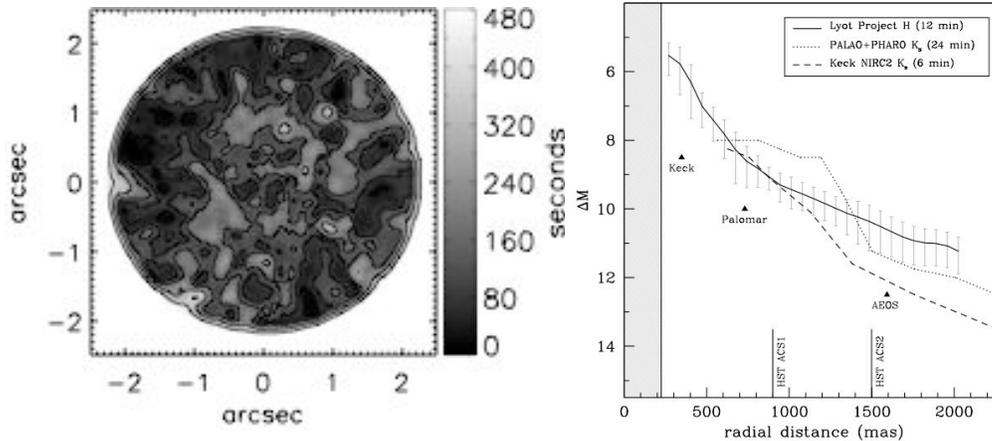

**Figure 11.** Left: Map of speckle lifetimes derived from hundreds of sequential images through a coronagraph behind an AO system (from Hinkley et al. 2007). Right: Dynamic range curve for a variety of systems showing that the speckle noise limit is essentially independent of telescope diameter. The systems as listed in the legend acquired data on 3.6-, 5- and 10-m telescopes. The labeled triangles indicate the AO "control radii" for each of the systems (see §6.4), and for comparison, the HST coronagraphic blocking spot radii are indicated along the bottom axis.

Speckles are also present even in space-based observations, where high Strehl ratios are achieved because the turbulent atmosphere does not corrupt the incoming wave front. Speckles in these systems, due to optical errors and minute fluctuations in the optics, also change with time and space craft conditions, although the timescales for these changes are much longer than for ground-based observatories. If the space-based systems did not have changing speckles, PSF subtraction would work at all radii from a given image of a star, because the speckle pattern would be determinable to a very high precision with deep observations of many stars. HST PSF subtraction attempts clearly show that this is not the case (Figure 2, e.g).

Several authors and teams building the next generation of high-contrast imaging devices have attempted to place a quantitative constraint on how well wave fronts must be controlled for a given contrast performance. In Table 2 we reproduce one such table based on the elaborate and comprehensive work of Stapelfeldt et al. (2006). Other studies (Macintosh et al. 2006, Dohlen et al. 2006, Levine, Shaklan & Kasting 2006) are essentially consistent with this. At this point, the reader should only refer to the first four columns. The rest of the table will be discussed in the context of whether there are fundamental limits to the contrast achievable from ground based telescopes, therefore requiring extraterrestrial platforms, in §5.

In summary, speckles are bright---roughly 250 times fainter than the primary star at a separation of about 7λ/D in typical modern instruments. They are correlated, and "blurring them out" with long integration times or by using broadband observations over large wavelength ranges, results in no increase in sensitivity to objects fainter than the speckle background.

### 4.3. Adaptive Optics

The effort to achieve diffraction-limited images on telescopes is generally referred to as adaptive optics, a broad field of research with applications outside of astronomy as well. Our treatment of this subject is necessarily minimal here, and we refer the reader to several excellent overviews, some of which tie directly into the primary subject of this article (Beckers 1993, Hardy 1998, Tyson 2000, Roddier 2004, Duchene 2008). Here, we simply present the quantitative limitations of AO with respect to high-constrat observations.

The first step in AO is achieved with a tip/tilt system, or fine guidance tracker, the lowest-order correction possible (Noll 1976). This system corrects for large movements (up to a few arcseconds) of the stellar PSF due either to atmospheric variations, wind, or vibration in the telescope. Once the image of the star has been stabilized by the tip/tilt system, the remaining correction to the wave front is achieved with a deformable mirror. These mirrors typically have hundreds (Roberts & Neyman 2002) or thousands (Dekany et al. 2006) of actuators underneath a thin reflective layer and can correct up to a few microns of wave front phase error. The shape applied to this deformable mirror is determined with a wave front sensor at rates of up to a few thousand times per second. Such sensors require sufficient light from a "guide star" to retrieve the wave front shape, and as a result, the faster the system operates and the finer the wave front control, the brighter the star has to be. (Table 2 columns 5 and 6 give approximate timescales and guide star brightnesses required of a large ground-based telescope to achieve the contrasts listed.) The goal of any adaptive optics systems is to create as flat a wave front as possible resulting in a high Strehl ratio image, with a PSF core approaching the diffraction limit.

Several surveys for companions to nearby stars using AO alone have proven to be productive, but they generally do not operate in the high-contrast regime (Chauvin et al. 2002, Neuhauser et al. 2003, Brandeker et al. 2003, Beuzit et al. 2004, Masciadri et al. 2005, Lafreniere et al. 2007, Nielsen et al. 2008). For example, these surveys have discovered several very young companions of stars or brown dwarfs which may be very low mass, but whose nature lying at the brown dwarf/planet boundary has been extensively debated: 2MASSWJ 1207334-393254 B (Chauvin et al. 2005a); GQ Lup B (Neuhauser et al. 2005); and AB Pic B (Chauvin et al. 2005b).

The primary purpose of AO in high-contrast imaging has been the production of diffraction-limited images, to which diffracted light suppression techniques can be applied (§5). Most AO systems, however, only operate at Strehl ratios of around 20-60%, and the resultant data produced by, for example, a coronagraph or interferometer behind an AO system, is generally limited by the remnant speckle noise. There are numerous reasons for this primarily due to specific instrument designs, but Table 2 presents physical limits to AO correction, as a function of required contrast. The contrast dictates the quality of the wave front necessary, and from this, one can estimate, based on an atmosphere model, the number of actuators needed on the deformable mirror and how quickly the AO system must operate its control loop (second-to-last column Table2). Another way to think about this is that a given wave front control requirement dictates the update rate for the AO system, say one with $10^4$ actuators on a 30-m telescope, and thus the faintest star that could be used as a guide star (given that one needs to be able to sense the wave front in each subaperture of the telescope). These quantities (Table 2, adapted from Stapelfeldt 2006) show that AO alone cannot solve the problem of high contrast observations. Indeed, to reach a contrast of $10^{10}$, to study evolved rocky planets, for example, requires a guide star of -8$^m$ in the H-band. Such stars do not exist.

AO can also be used to control speckle noise, but before dealing with that, we describe the techniques that control diffracted light.

# 5. SUPPRESSION OF DIFFRACTED LIGHT
## 5.1. Classical Lyot Coronagraphy

Here we briefly summarize the idea behind the Lyot coronagraph, as first described by Lyot (1939) and later quantified and optimized for application to high-contrast imaging by Sivaramakrishnan et al. (2001). The coronagraph is shown schematically in Figure 12, for the case of no wave front errors. It uses two masks to achieve the suppression of starlight. In the first stage, an image of the target star is formed at the center of a circular, opaque focal plane mask. Optimally such masks should have diameters of about 3-6 $\lambda/D$, meaning that light from an object 1.5 to 3 $\lambda/D$ radially away from the star is not occulted by this mask. The starlight is largely absorbed by this mask, but also diffracts around it. The beam after the focal plane mask is then brought back out of focus and an image of the telescope pupil is formed. In this plane, much of the residual starlight has been concentrated into a bright outer and inner ring around the conjugate location of the secondary telescope mirror (if there is one). This concentration of the starlight is critical to understanding a coronagraph and is due to the diffraction caused by the focal plane mask. Any object's light that is not significantly diffracted by the focal plane mask will be distributed in this pupil image evenly. In this manner, the coronagraph has effectively separated the primary star's light, by using diffraction, away from that of a faint object next to the star. It can thus be filtered out further without greatly affecting the light from the fainter object. Indeed, the eponymous Lyot mask is placed in this pupil image. It downsizes the telescope aperture slightly while slightly increasing the size of the secondary obscuration, simply in order to block the bright concentrated rings of the central star's light. Finally, optics form an image after this Lyot mask, where the overall intensity of the central star has been reduced by more than 99%, while a neighboring object will only be affected at the few percent level.

There are numerous coronagraphic surveys that have been completed or are still underway at Palomar (Oppenheimer et al. 2001, Metchev & Hillenbrand 2008, Hinkley et al. 2008, Dekany et al. 2007, Serabyn et al. 2007), Subaru (Fukagawa et al. 2004, Itoh et al. 2006), AEOS (Oppenheimer et al. 2003, 2004), VLT (Beuzit et al. 2006, Dohlen et al. 2006), Gemini (Macinosh et al. 2006, Artigau et al. 2008), MMT (Biller et al. 2007) and the William Herschel Telecope (Thompson et al. 2005).

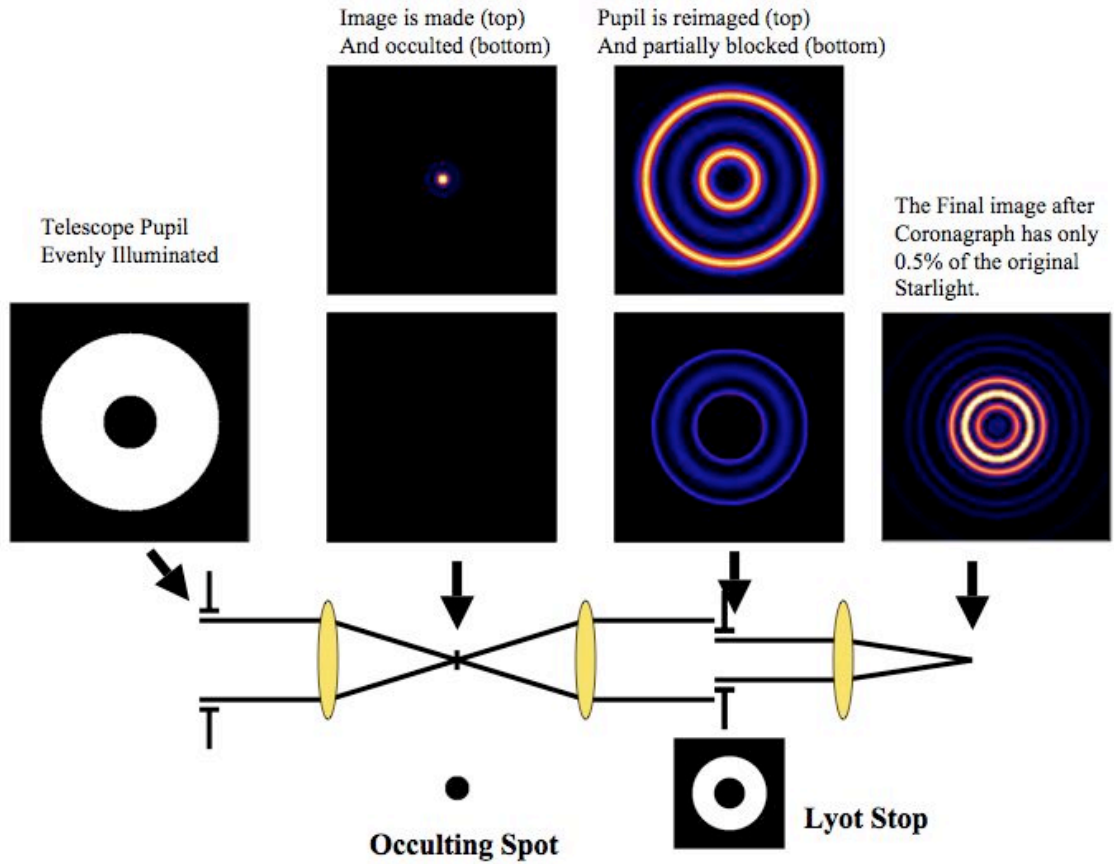

**Figure 12.** Basic theory of Lyot Coronagraphy, adapted from Sivaramakrishnan et al. (2001), and assuming no wave front errors throughout the system. Along the top images show the distribution of intensity at each critical point in the simplified optical train shown along the bottom.

### 5.2. Apodized Coronagraphs

The classical system described above can be improved in many ways, but the most practical and simplest way is by attempting to "soften" the diffractive effects of the telescope and focal plane mask. This can be done by placing a graded, or apodized, transmission function on the telescope pupil prior to the focal plane mask. The efficacy of this apodization for causing the image at the focal plane mask to be far more concentrated, with a much faster radial fall-off has been known for some time (e.g. Jacquinot & Roisin-Dossier (1964), but the application to coronagraphy was fully elucidated by Soummer (2003, 2005), who found transmission functions based on prolate spheroidal functions that optimize combinations of apodizers, focal plane mask sizes and Lyot masks. In fact in the ideal case, the Lyot mask is no longer necessary in these systems. The effect of apodization is actually two-fold: it reduces the brightness of the Airy rings and also reduces the contribution to the overall noise budget from speckles which are "pinned" to the Airy rings (Bloemhof et al. 2001, Sivaramakrishnan et al. 2002). Such speckles tend to be brighter than most, and as such should be reduced first.

Phase-Induced Amplitude Apodization, is, at its core an identical idea to the apodized coronagraph, but it has the benefit of acheiving the goals without any loss of throughput (Guyon et al. 2003, 2005). This is done with a relatively complicated set of optics, usually mirrors with soft, multi-element curvatures deviating from a standard parabola that redistribute the intensity in the conjugate pupil plane by changing the propagation direction of light rays through the system. This allows all the light to propagate through the coronagraph, rather than simply absorbing some of it as in the apodized coronagraph. Only after the focal plane suppression are the rays put back in their normal angles, so that image quality is maintained in the final coronagraphic image. Initial results on the sky have been obtained for this technique, and it appears promising, although manufacture of the optics is difficult and relatively expensive (e.g. Kenworthy et al. 2007).

### 5.3. Non-Classical "Coronagraphs"

The class of techniques that fall under the appelation "coronagraph" at this point is very wide with no less than twenty different ideas on how best to suppress starlight in a classical imaging device (a single telescope). The ideas are the subject of a rich literature, with various different studies of their comparative strengths and weaknesses. Since most of these ideas have not actually been used to observe stars yet, we refer the reader to a series of articles that compare and contrast these techniques in great detail from a theoretical perspective, and describe a few of them with particular promise. Perhaps the most comprehensive treatment of the relative merits of each technique can be found in Guyon et al. (2006), in which 16 different techniques are compared. Ones of particular promise are the phase mask technique (e.g. Boccaletti et al. 2004, Palacios et al. 2005), shaped pupils (e.g. Kasdin, Vanderbei & Belikov 2007), occulters (e.g. Cash et al. 2007) and band-limited masking (e.g. Kuchner & Traub 2002). Generally speaking, the precise control of diffracted light with these techniques assumes that optical wave front errors are not present in the real system. As such most of these non-classical techniques are presently only applicable to space-based observations, and even then their successful implementation will require AO as well, though with less stringent requirements.

#### 5.3.1. Band-Limited Coronagraphs

The band-limited coronagraph (Kuchner & Traub 2002) was conceived as a perfect coronagraph which would entirely eliminate the central star's PSF in the absence of any wave front error. The fundamental principle of band-limited focal plane mask coronagraphy is tied up in the notion that a focal plane mask could be shaped in such a manner that any residual starlight diffracted by that mask would be thrown entirely into a highly localized region of the subsequent re-imaged pupil, and thus easily rejected with a Lyot mask that only masks these very specific regions.

In general, the transmission profile of a focal plane mask can be written as $w(\eta,\xi) \equiv 1 - m(\eta,\xi)$. The arguments are coordinates in image space. The band-limited design makes use of a focal plane mask function $m(\eta,\xi)$ that only has certain spatial frequencies represented. Its Fourier transform $M(x,y)$ is non-zero over only a finite area of its domain. Thus it is band-limited with $M(x,y) = 0$ if $|x^2 + y^2| > b^2$. In the parlance of Fourier theory, $b$ is the bandpass of $m$, even though $b$ is actually a physical distance in pupil space, but corresponding to a spatial frequency in the image plane. To design a band-limited coronagraph, the occulting mask function $m$ is selected based on its properties in its transform (pupil) space, $(x,y)$, rather than in physical (image) space, $(\eta,\xi)$. In other words, one chooses a pupil after the occulting mask that has exactly zero light over the great majority of the telescope entrance pupil. This can be understood as simply an inverted "top-hat" function, allowing light in only one location in the telescope pupil but no where else.  The Fourier transform of this fucntion yields the actual focal-plane occulting mask shape.  The Fourier transform of the top-hat function *M* is just a two-dimensional *sinc* function, and the size of this stop is defined by the bandpass of the mask function in spatial frequency space, or in the pupil plane.  If the telescope diameter is *D*, then the characteristic scale of the mask function is *D/b* resolution elements (which are themselves $\lambda/D$ in angular extent, where $\lambda$ is the wavelength of light being considered). Thus the final mask will be approximately *D/b* Airy rings in width at the image plane.

In spite of the obvious theoretical advantages of the band-limited coronagraph, the design works ideally only at one wavelength.  Attempts to make it work over a broad wavelength range result in very significant reductions in overall throughput of the system, especially in systems with secondary mirrors and spider supports in the telescope pupil.

### 5.3.2. Shaped-Pupil Coronagraphs

Shaped-pupil "coronagraphs" rely on the principle of cutting the incoming wave front in the pupil plane, into a series of optimized shapes to produce a PSF with the contrast required to image and characterize a planet.  This is different from a traditional Lyot coronagraph design, which requires an occulting stop in the image plane to block the on-axis stellar PSF.  In this case, functions that produce diffraction patterns in the image plane can be solved for with particularly dark regions around the central star.  An example of one for an 8-m telescope similar to the Gemini pupil geometry is shown in Figure 13**.**  This design has modest throughput of about 30%**.**  The resultant PSF has bright and dark wedges within which objects as faint as $10^{-7}$ contrast could be detected. An observing strategy using this pupil would require at least three observations of each star to enable point-source detection at any position angle relative to the bright AO target. While shaped-pupil designs work extremely well for systems without residual wave front error, recent numerical and theoretical work has shown their performance degrades substantially for non-perfect wave fronts (in particular, see the conceptual design documentation for the Gemini Planet Imager, available from the Gemini Observatories), meaning that they are not an attractive option at Strehl ratios of less than ~98% (Figure 13).

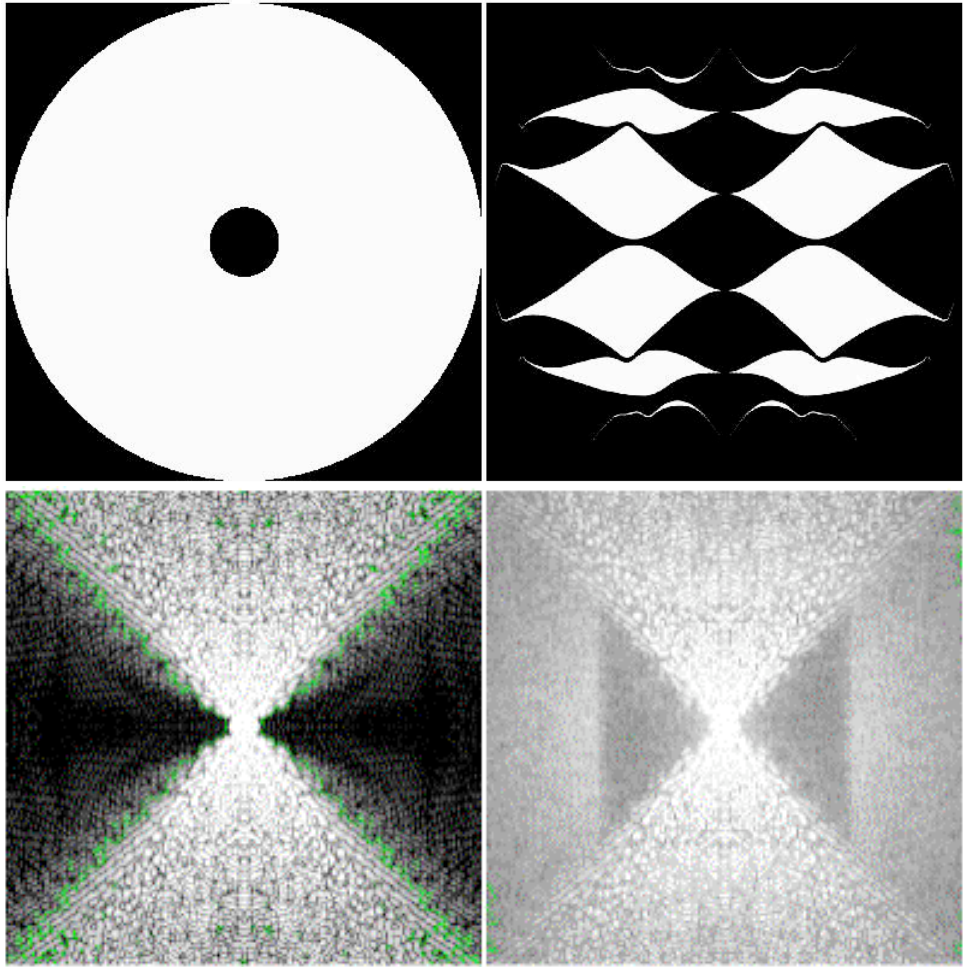

**Figure 13.** Shaped Pupil coronagraphy example. A standard pupil illumination pattern with a secondary obscuration (top-left) and an optimized shaped pupil mask (top-right) that generates the bottom-left PSF in the corresponding pupil plane in the assumption of perfect optics. The darkest regions of that PSF are $10^{-7}$ as bright as the core of the central star (green contour). The bottom-right shows the same PSF with a 96% Strehl ratio AO-corrected wave front. That PSF is brighter than $10^{-6}$. Everywhere in the 4 arcsecond simulated image (simulations courtesy of A. Sivaramakrishnan, J. Kasdin).

Another problem with shaped pupil systems, even if they can be achieved with near-perfect wave fronts, is that the complicated, extended PSFs prohibit studies of extended objects around target stars. Objects such as circumstellar disks or even multiple point sources in the field of view will result in the convolution of this PSF with the real infinitely resolved on-sky intensity distribution. The bright parts of the bow-tie structure shown in Figure 13, or rings in a symmetric function solution, will simply fill the field of view, reducing contrast substantially.

### 5.3.3. Phase-Masking

Phase Mask coronagraphs (e.g. Boccaletti et al. 2004) modify the wave front at the image plane by imparting phase differences between different parts of the image to create destructive interference of on-axis light. In theory this technique could remove all of the starlight while having no effect on anything else in the field of view. Several versions of this type of coronagraphy exist with various types of geometry for the phase mask, with some using four quadrants, some round regions and more recently tilted phase inductions called "optical vortices" (e.g. Rouan et al. 2000, Soummer et al. 2003, Palacios et al. 2005, Swartzlander et al. 2008).

The simplest phase mask to understand is the four-quadrant one. In this case, an optic is inserted into the focal plane in which two quadrants of the image on a diagonal have their phase retarded by $\pi$ radians. In the subsequent conjugate pupil plane, these four quadrants, then, will recombine in destructive interference. However, this works at only one wavelength. According to Rouan et al. (2000), the rejection rate ($r$) for a Four Quadrant Phase Mask (FQPM) coronagraph as a function of wavelength $\lambda$ is given by

$$r(\lambda) = \frac{4}{\pi^2}\left(\frac{\lambda}{\lambda - \lambda_0}\right)^2,$$

where $r$ is the ratio of the unmasked PSF to the phase mask image. For the case where $\lambda = \lambda_0$, the rejection is infinite. However, even for small differences between $\lambda$ and $\lambda_0$, the rejection rapidly decreases. Thus, phase masks are, by nature, finely tuned to a specific wavelength of interest. Achromatic phase mask designs using zeroth order gratings or multi-layer depositions may mitigate some of the chromaticity problems, and numerous experiments are ongoing. The latest optical vortex results indicate starlight suppression in real telescopes and laboratory experiments on the order of a factor of 100 (Swartzlander et al. 2008). These sorts of techniques generally require extremely precise pointing, since a displacement of the star from the center of the phase mask greatly reduces its efficacy.

### 5.3.4. Occulters

Recently an idea came into prominence which involves an optical stop positioned at a tremendous distance from the telescope used to conduct the observations. Cash et al. (2007) proposed placing a large star shaped mask many tens of thousands of kilometers in front of a telescope in the line of sight to a distant star around which one wants to detect objects at the $10^{-10}$ level or fainter. The idea behind this is that this star shaped mask, generally much larger than the telescope aperture itself (roughly 50 m for a 6 m telescope), but smaller in angular extent at that distance than the region of interest around the star, causes diffraction of the starlight such that it never enters the telescope in the first place. Generally an occulter of this nature would not create a proper shadow on the telescope, because the diffraction at such a huge distance creates the so-called spot of Arago, a bright spot that would preclude imaging any high-contrast target. However, the occulter shape can be optimized much like the optimizations of the complex shaped-pupil masks to permit the occulter idea to work. These shapes can also be modified to allow the occulter to work over large wavelength ranges (e.g. Cady et al. 2007).

The fact that the starlight does not enter the telescope in the first place greatly reduces many of the problems described in this article, and this has become the subject of a number of studies for space missions where the occulter could be placed at such huge distances from the observatory. Some drawbacks are that the occulter has to be repositioned over vast distances to observe another star, it must be built and deployed to demanding mechanical precision, and it must be kept in place relative to the star and the observatory to within a meter or less.

In all cases of "coronagraphs," each technique has drawbacks. As we showed, some are only applicable to imaging of point sources at high contrast, but would be utterly insensitive to diffuse illumination through the field of view. Some have extremely limited fields of view and some have stringent requirements on wave front errors that go beyond the values in Table 2. All in all, classical coronagraphs, with the improvements due to apodization, as described above, have been by far the most productive and most mature of the various techniques. We also refer the reader to the review of Beuzit et al. (2007), Coulter's (2005) volume, as well as the volume by Aime & Vakili (2006), which contains many articles on specific techniques with very recent results.

### 5.4. Coronagraphs and Astrometry

There are several challenges to working with coronagraphic data. Most notably, the precise position of the star behind the coronagraphic mask is difficult to measure. The point, after all, is to get rid of the star. However, such a measurement is extremely important for many of the scientific results in §2. For example, the relative position between the host star and any companions is required to establish physical association and to study orbital motion. A valuable solution to this problem was put forth by Marois et al. (2006b) and Sivaramakrishnan & Oppenheimer (2006), in which a periodic grid or a sinusoidal ripple on the wave front is inserted at the telescope pupil. This causes four fiducial images of the occulted star to appear at known locations relative to the star outside the coronagraphic mask. The intersection of the two lines specified by spots on opposite sides of the star determines the star position with an accuracy that can be chosen for a given system or scientific goal. This technique is essentially the intentional introduction of permanent, well-understood speckles into the image that precisely locate the star. In addition, these calibrator spots also have a known brightness, based on the design of the grid, allowing accurate relative photometry between an occulted and an unocculted object within the coronagraphic image.

### 5.5. Interferometers

Optical and IR interferometry has matured into an extremely important part of the astronomer's arsenal of observational techniques, wonderfully reviewed, as applied to the science of circumstellar material, by Akeson (2008). In the realm of high-contrast observations, however, the technology is far less mature than the class of coronagraphic techniques we mentioned above. This area, though, has tremendous potential. In principle one can use an interferometer to null light from a bright star such that objects in close proximity are detectable. The major advantage that interferometers possess is a far higher angular resolution than single telescope imaging programs can achieve. As such it is interferometers that may eventually be the standard bearer for comparative exoplanetary science. Consider, first, the fact that a system such as the CHARA array on Mt. Wilson, with a maximum baseline of some 300 m can resolve even some of the hot Jupiter planets from their primary stars (ten Brummelaar et al. 2008, Beuzit et al. 2007, and references therein). Indeed, the contrasts for such objects, as shown in the Swain et al. 2008 work are at the level of $10^4$.

One type of measurement with an interferometer is direct nulling, which has the potential to obtain spectra of planets or other faint compainions if calibrations and systematic errors can be understood precisely. A nulling interferometer (e.g. Serabyn et al. 2000) inserts an extra phase delay in one arm of the interferometer so that light from the central star destructively interferes. Since the null depends on the incident angle of the starlight, the interferometer can be pointed carefully, such that it only negates the star's light while allowing light from the planet to reach the final detector.

Nulling interferometry has been achieved at a contrast of about $10^2$ on the Keck interferometer system (G. Vasisht, personal communication), and at the MMT system (Hinz et al. 1998). It is extraordinarily difficult to achieve better results from the ground because of the deleterious effects of the turbulent atmosphere. Once again, as with the coronagraphs, extremely well-controlled input wave fronts of light are required, and in the interferometer the separate paths of light must be matched in length to better than a nanometer to reach the $10^5$ or better contrast level. These are mitigated partially by using longer wavelengths of light.

The primary difficulty, once these calibrations are understood, will be system optical throughput. While interferometers can achieve unprecedented resolution, the collecting area is generally quite small, and extremely long integration times are required for direct observation of exoplanets at contrasts above $10^5$. That said, an enormous amount of work has gone into defining sophisticated space-based interferometers capable of detecting even Earth-analog planets within some 10 pcs of the Sun (e.g., Fridlund 2008 and references therein, Lawson et al. 2006). A more detailed review of interferometry for high-contrast observations can be found in Beuzit et al. (2007).

## 6. SPECKLE SUPPRESSION

As discussed in previous sections, the primary difficulty that high-contrast observing must overcome is uncontrolled and unmeasured wave front error, typically with an rms value of less than $\lambda/10$. At this level the Strehl ratio is high, image core stability is superb, and image quality is better than required for most observational efforts in astronomy. However, it is insufficient to detect objects $10^5$ times fainter than another object within a small angle ($< 40\ \lambda/D$). Table 2 shows an estimate of the wave front quality needed to achieve various contrast levels in the presence of a device that achieves the suppression of the diffracted light (e.g. an apodized Lyot coronagraph). Note that even for the lowest contrast in the table, the wave front must be controlled at the $\lambda/100$ level, beyond the performance of existing AO systems in operation on the sky today.

Clearly the primary difficulty is in controlling the wave front, and suppressing speckles in the image plane, not in achieving the suppression of the diffracted light pattern. It should be noted here, though, that sufficiently good ideas in the diffraction arena could possibly eliminate the need for such precise wave front control. Those claiming this have not convincingly demonstrated this in laboratory experiments so far, but this is a partial explanation for the abundance of these new types of coronagraphs, and the frequency with which papers announcing new techniques appear in the literature. We also note that although interferometers and coronagraphs operate in fundamentally different manners, both are stymied by this same fundamental problem of residual optical error, but in interferometry one deals with that phase error directly rather than with speckles.

There are two approaches to mitigating these small wave front errors: control them or remove their manifestation, speckles, through special data collection and processing techniques. Even better, one could use both approaches, and several new systems are planning to do this. Far more effort has gone into speckle suppression in images than the equivalent in interferometry (more accurately referred to as calibration), so we emphasize imaging in this section.

In the case where one attempts to remove or, more appropriately, suppress speckles (as opposed to controlling them), data must be acquired that is sensitive to properties of speckles that are not shared with real celestial sources. This permits the data to be processed so that the speckles can be distinguished from real sources and removed. As we described in §4.2, speckles appear at locations in the focal plane that are a function of (1) time, (2) wavelength, and (3) the orientation of the sources of wave front error that cause them. Any real source will not move around in the image plane significantly (assuming the system is operating properly at S > 10 % and has no major optical design flaws) as a function of any of these three parameters, unless it is either extremely close to Earth (i.e. an asteroid), or has a tremendous proper motion for another reason. (Note that the existing interferometers can detect the proper motion of Barnard's star in a matter of tens of minutes.) A fourth quality of speckles that is different from some types of sources is that they are generally not polarized, because starlight is generally not polarized. Thus if one is searching for a source exhibiting polarization, speckles could be suppressed using Stokes vector measurements. In Figure 14 we present a dynamic range plot (see Figure 11) showing the effect of several speckle suppression techniques.

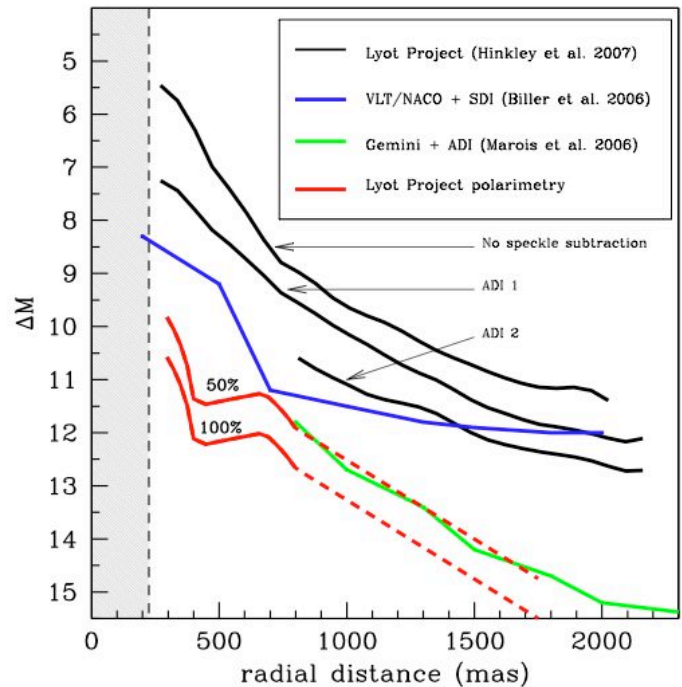

**Figure 14.** Dynamic range curves for a variety of speckle suppression techniques. The upper most curve is the same as the one in Figure 11, and all curves are given as 5-σ detection limits. The two black ones below it are examples of removing rotating patterns of speckles with the ADI (§6.1) technique on two separate rotations for that Lyot Project data. Curves for SDI (§6.3) and ADI from other systems are show as well. The two red curves for Double-Difference Polarimetry (§6.2) come from Oppenheimer et al. (2008). These represent detection limits for 50% and 100% polarized sources and include an extrapolation of the slope (red-dashed part of the curves) because the system's limiting magnitude was reached in those observations. Plotting such horizontal lines on this graph would have detracted from the comparative purpose, but the dashed lines are an appropriate estimate of the double-differenced PSF considering other Lyot project data.

### 6.1. Angular Differential Imaging

One technique, called "angular differential imaging" (ADI) takes advantage of rotation of speckles with respect to the sky, if observations are taken on non-equatorially mounted telescopes. In the case of alt-az telescopes, an instrument at the Cassegrain focus will see speckles rotate as a star is observed, because the primary mirror rotates with respect to the sky. An instrument at the Coude focus of such a telescope will see at least two different rotations (e.g. Figure 14, Lyot Project "ADI 1" and "ADI 2" curves, Hinkley et al. 2008). Marois et al. (2006a) showed that due to the quasi-static nature of the speckle pattern due to these rotating optics, some of the speckles can be suppressed through post-processing given a sufficient angle of rotation, with many images taken during that rotation. ADI requires complex data processing, and can only work on telescopes where rotations exist, but it can provide one to two magnitudes of improvement, and, at least in the case of Marois et al. (2006a), three to four magnitudes. Artigau et al. (2008) are combining ADI and SDI (See §6.3) simultaneously in an experiment that promises interesting results.

### 6.2. Polarimetry

Double-differential polarimetric imaging can very effectively suppress speckles leaving any polarized structures behind. This requires taking images in different polarization states simultaneously, as described by Kuhn et al. (2001), who successfully imaged circumstellar disks, in which dust grains scatter starlight, inducing polarization. Perrin et al. (2004) used the technique to image the dusty regions surrounding Herbig Ae/Be stars using laser guide star AO, and most recently, Oppenheimer et al. (2008) used this technique to obtain an image of the disk surrounding AB Aurigae at a contrast of $\sim 10^5$ as discussed in §2.

In Figure 14 we show that polarimetry is one of the most effective speckle suppression techniques. The curves shown are based on data from the Lyot Project (Oppenheimer et al. 2008), and include an extrapolation (dashed portion of the curve) based on the slope of the double-differenced PSF. Unfortunately the instrument's background noise floor was reached at the point where these dashed curves begin. The images were limited by detector sensitivity rather than residual starlight. However, based on the other curves, and the data the extrapolation is a very likely representation of the actual double-differenced polarimetric PSF. We went through this exercise here because Figure 14 is meant to represent the various techniques, not to show a particular instrument's sensitivity.

### 6.3. Exploiting the Chromatic Behavior of Speckles

Racine et al. (1999) and other authors suggested subtracting two images of a star at two closely separated wavelengths across the methane band head (1.59 microns), thus removing the stellar halo, but revealing any cool methane containing companions. One chooses wavelengths very close together, so the PSF is as similar as possible in the two images, and must attempt to measure the two images simultaneously or the speckle pattern will change from one to the next. This technique with three simultaneous images was the primary objective of the TRIDENT instrument (Marois et al. 2005) as well as another experiment by Biller et al. (2006). Referred to as "simultaneous differential imaging" (SDI), the method produces gains in dynamic range of about one or two magnitudes (Figure 14).

Increasing the spectral resolution can vastly increase the power of this technique, and it becomes a somewhat different speckle suppression method, initially suggested by Sparks & Ford (2002). The speckle noise pattern is chromatic, with the speckles moving radially in an image as a function of wavelength (see radial pattern in Figure 12). Therefore, data from an integral-field spectrograph, or a hyperspectral imager, where many images at a range of wavelengths are obtained could be used to remove speckles. In this type of data, the speckles follow diagonal paths through a data cube, while any genuine astrophysical structures have fixed positions with wavelength. The Gemini Planet Imager, Project 1640 and SPHERE are all employing this type of speckle suppression (e.g. Macintosh et al. 2006, Hinkley et al. 2008, Dohlen et al. 2006). This type of data also allows for extraction of spectra of the objects of interest with a resolution of between 30 and 80.

### 6.4. Speckle Control

In addition to the speckle suppression techniques, it is also possible to control both the PSF and the speckles with an AO system and specialized wave front sensors. An AO system has a "control radius" ($\theta_{AO}$; Malbet, Yu & Shao 1995) defining a usually square (depending on the geometry of actuators on the deformable mirror) region of a PSF which the AO system can influence, $\theta_{AO} = N_{act} \lambda/2D$, where $N_{act}$ is the linear number of actuators across the deformable mirror. The simplest way to understand the control radius is that the deformable mirror can only influence spatial frequencies in the pupil plane at the Nyquist frequency $k_{AO} = N_{act}/2D$. The image plane is a Fourier transform of this pupil plane so that spatial frequency is transmuted into an angle, within which the AO system can control PSF shape and speckles. Outside this radius, the AO system has no control over the PSF.

The AO system, therefore, can effectively permit engineering of dark features in PSFs, and can be used to remove a speckle. A speckle can be deleted by determining the sinusoidal ripple in the wave front that it is due to and imposing the opposite ripple on the deformable mirror. This technique is described in Trauger & Traub (2007), and especially in Wallace et al. (2004) as well as others. It is an iterative technique and requires an extremely sensitive wave front sensor, generally working much slower than the usual AO wave front sensor. For example in the Gemini Planet Imager, this is achieved with a Mach-Zender interferometer, measuring the speckles on a 1 s timescale to correct the longer lived ones at a wave front accuracy of about 1 nm or $\lambda/1000$ at the science operating wavelength (Macintosh et al. 2006). Both SPHERE and Project 1640 have similar systems as well (Dohlen et al. 2006, Hinkley et al. 2008). All three of these systems and a similar one for the Subaru Telescope (Hodapp et al. 2006) are attempting to reach contrasts of $10^6$ to $10^8$ within 10 $\lambda/D$ of the brightest nearby stars by early 2011. These are inherently complex instruments, and are precursors to similar systems envisioned for the next generation of extremely large telescopes with apertures of around 30 m. For example, the opto-mechanical layout for only the adaptive optics system of the Gemini Planet Imager project is shown in Figure 15. This system also includes a cryogenic hyperspectral imager, a coronagraph, and two wave front sensors (none of which are shown). Kenworthy et al. (2006) demonstrated the speckle control techniques on an actual telescope rather than just in the laboratory.

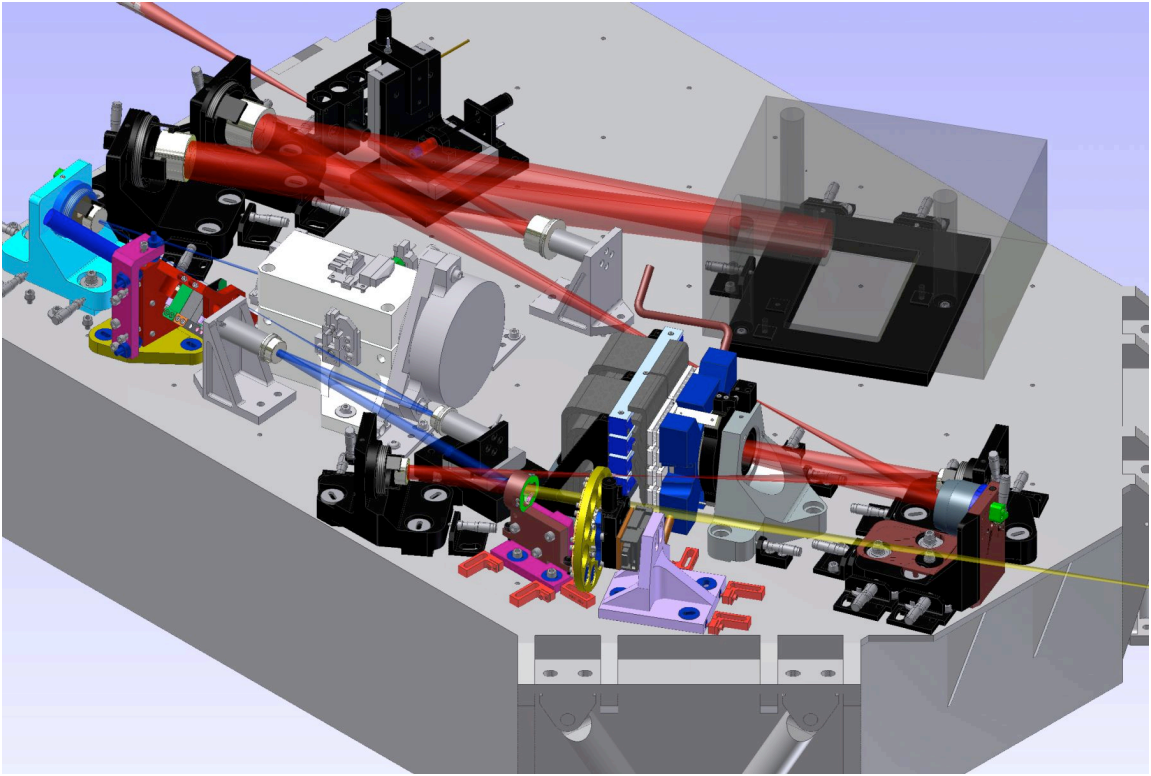

**Figure 15**. The Gemini Planet Imager's adaptive optics relay, shown to demonstrate the complexity of the next generation of planet imaging devices. This system includes approximately 3300 moving parts, and the hyperspectral imager is not shown. (Courtesy Darren Erickson).

## 7. IS THERE A LIMIT TO HIGH-CONTRAST OBSERVATIONS?

A number of laboratory experiments, primarily in support of the Terrestrial Planet Finder (TPF) project and DARWIN have demonstrated in laboratory environments that contrasts of $10^{10}$ can be achieved in a relatively short period of time with speckle control algorithms similar to those mentioned above in a space-based environment. These demonstrations are critical, first for validating the effort going into the systems being built now, and, second, to show that far more advanced systems such as DARWIN and TPF are not only feasible, but have practical solutions already (Trauger & Traub 2007, e.g.). Given the immensely compelling nature of the science involved in detecting places that might host life outside our solar system, there is no question that, barring the annihilation of homo sapiens, people will, and in some sense must, conduct such missions. We know how to do them now.

However, probably before such missions happen, another round of high-contrast work on ground based telescopes will occur after the advanced AO coronagraphs on Gemini, Subaru, VLT and Palomar have finished their surveys. These will be the classes of experiments on the upcoming 30-m scale optical and IR telescopes. Various authors have written about and have advanced the designs of such systems. In principle these should be able to reach the contrast level of $10^8$ at far closer separations than ever before

36                                                                                                                    Oppenheimer and Hinkley

achieved and with much greater sensitivity. Why only $10^8$, when TPF (which likely will be a much smaller telescope) will reach $10^{10 \text{ to } 12}$? First of all, we showed above that contrast is not a strong function, if any, of telescope aperture, and we emphasized the extreme importance of wave front control. But is there a fundamental limit to work on the ground or even in space?

Table 2 provides one perspective on this. As Stapelfeldt (2006) points out, when considering the control of speckles, the level of wave front control required for a $10^4$ actuator system on a 30-m telescope to reach a contrast of $10^8$ requires measurements from the wave front sensor so rapidly that one could only look at the few brightest stars in the sky to reach such a contrast. Reaching $10^{10}$ would simply be impossible. Thus, the argument goes that anything beyond $10^8$ must be done from space, where wave front errors evolve on far longer time scales, and such levels of control could be conducted on far fainter target stars as the AO system reference point source.

On the one hand this makes perfectly logical sense. On the other hand, if speckle suppression, rather than control, in all its manifestations (along with new ideas for post-processing removal of speckles from images) continues to perform at the 2-5 or 6 magnitude improvement (factors of 10 to a few hundred), one might be far more optimistic about ground based projects. In any case, such questions will be answered very soon with real data. We can also remain optimistic about these technique issues, because many people are rethinking basic ideas and assumptions that go into the concepts for new techniques. Perhaps someone will produce an observation technique that is not limited by diffraction, for example, or that naturally filters out the wave front errors and greatly reduces the engineering requirements on these systems.

## ACKNOWLEDGEMENTS

The authors thank the following for contributions to this article in the form of figures or comments on the text or lively discussions: Charles Beichman, Richard Dekany, Karl Stapelfeldt, Anand Sivaramakrishnan, Michael Shara, Remi Soummer, Gautam Vasisht, Isabelle Baraffe, Gilles Chabrier, Matthew Kenworthy, Ian Parry, Bruce Macintosh, Les Saddlemyer, Darren Erickson, Alicia Stevens, and Futdi.

Table 1. Comparison of Contrast and Resolution Requirements for Science Goals

| Science Goal | Physical Scale | Angular Resolution (mas) | Resolution Elements[3] ($\lambda/D$) | Contrast for Detection | Contrast for Spectroscopic Study |
|---|---|---|---|---|---|
| Sun's Corona[1] | <0.001 AU | >1000 | >40 | $10^6$ | $10^{6 \text{ to } 9}$ |
| Quasar Host Galaxies[2] | 1-10 kpc | 40-1000 (z ~ 0.4) | ~2 to 40 | 10-100 | $10^3$ |
| Young (<1 Gyr) Jupiter-Mass exoplanets | 1-50 AU | 100-5000 | 4-200 | $10^6$ | $10^8$ |
| Old Jupiter-Mass exoplanets | 1-50 AU | 100-5000 | 4-200 | $10^8$ | $10^{8 \text{ to } 10}$ |
| Evolved Earth-mass planets | 0.1-10 AU | 10-1000 | <1 to 40 | $10^{10}$ | $10^{12 \text{ to } 14}$ |
| Evolved Star Outflows | 0.1-1000 AU | 10-$10^5$ | <1 to 40 | Unknown | Unknown |

[1] We do not consider observations of the sun's corona of particular technical interest in this article because generally observations of it do not meet all of the requirements of our definition of high-contrast. An eclispe or standard coronagraph are sufficient to achieve the science goals, and the angular scales are such that advanced optics techniques are not necessarily required to study the corona.

[2] Quasar host galaxies are generally detected at contrasts of 0.1 to 0.01, with the highest contrast example being just below $10^{-3}$ (Magain et al. 2005, Floyd et al. 2004).

[3] This column represents a simple conversion of the angular resolution value in mas to units of the diffraction-limited resolution (see definition box) of an 8-m telescope operating at an observing wavelength of $\lambda = 1\mu m$.

Table 2. Maximum Wave Front Error for a Given Contrast[1]

| Contrast | Coherent Wave Front Error[2] | RMS Wave Front Error[3] | RMS Path Length Error[4] | Reduced Coherence Time[5] | Guidestar H Magnitude[6] |
|---|---|---|---|---|---|
| $10^6$ | $\lambda/4,400$ | $\lambda/88$ | 18.7 nm | 1.07 msec | 6.9 |
| $10^7$ | $\lambda/14,000$ | $\lambda/280$ | 5.7 nm | 0.34 msec | 3.2 |
| $10^8$ | $\lambda/44,000$ | $\lambda/880$ | 1.9 nm | 0.11 msec | -0.6 |
| $10^9$ | $\lambda/140,000$ | $\lambda/2,800$ | 0.6 nm | 0.03 msec | -4.3 |
| $10^{10}$ | $\lambda/440,000$ | $\lambda/8,800$ | 0.2 nm | 0.01 msec | -8.1 |

[1]Reproduced with permission and adapted from Table 1 in Stapelfeldt (2006). The fifth and sixth columns are discussed in §5.

[2]Amplitude of a single sinusoidal phase ripple across the entrance pupil, whose corresponding focal plane image speckle would present the contrast to the central star specified in column 1.

[3]Reduced phase error after averaging over 2,500 incoherent modes. This corresponds to a $10^4$ actuator AO system, which, on a 30 m telescope, would yield a high contrast dark field 0.57 arcsec in radius at H band.

[4]For H-band ($\lambda = 1.65$ µm) science observations

[5]The time over which the phase of the atmospheric wavefront changes by less than the RMS error values given in columns 3 and 4, assuming a standard coherence time of 15 ms at H band.

[6]Guidestar brightness needed to completely sense the wavefront to the accuracy given in column 4, within the reduced coherence time of column 5, for $10^4$ subapertures on a 30 m telescope. A noiseless infrared interferometric wavefront sensor with 50% throughput operating at 20% bandwidth is assumed.